\documentclass[lettersize,journal]{IEEEtran}
\usepackage{amsmath,amsfonts}
\usepackage{algorithmic}
\usepackage{algorithm}
\usepackage{array}
\usepackage[caption=false,font=normalsize,labelfont=sf,textfont=sf]{subfig}
\usepackage{textcomp}
\usepackage{stfloats}
\usepackage{url}
\usepackage{verbatim}
\usepackage{graphicx}
\usepackage{cite}
\usepackage[T1]{fontenc}
\usepackage{multirow}
\usepackage{multicol}
\begin{document}

\title{NWaaS: A Non-Intrusive and Privacy-Preserving Watermarking-as-a-Service System with Adaptive Resource Scheduling}

\author{Haonan An, 
Qianyao Ren,
Guang Hua,~\IEEEmembership{Senior Member,~IEEE,}
Tao Li,
Yu Guo,
Yanan Ma,
Hangcheng Cao,\\
and Yuguang Fang,~\IEEEmembership{Fellow,~IEEE}
\IEEEcompsocitemizethanks{
    \IEEEcompsocthanksitem Haonan An, Qianyao Ren, Yu Guo, Yanan Ma, Hangcheng Cao, and Yuguang Fang are with Hong Kong JC STEM Lab of Smart City and Department of Computer Science, City University of Hong Kong, Hong Kong. \protect
    E-mail: haonanan2-c@my.cityu.edu.hk, qianyaren2-c@my.cityu.edu.hk, yu.guo@cityu.edu.hk, yananma8-c@my.cityu.edu.hk, hangccao@cityu.edu.hk, my.fang@cityu.edu.hk.
    \IEEEcompsocthanksitem Guang Hua is with the Infocomm Technology Cluster, Singapore Institute of Technology, Singapore 828608. \protect
    E-mail: ghua@ieee.org.
    \IEEEcompsocthanksitem Tao Li is with Department of Electrical and Electronic Engineering, University of Hong Kong, Hong Kong. \protect
    E-mail: litao@eee.hku.hk.
    }
\thanks{The work was supported in part by Singapore Ministry of Education (MOE) under the Academic Research Fund (AcRF) Tier 1 Grant R-MA123-R205-0008, by the National Research Foundation Singapore under the AI Singapore Programme (AISG Award No: AISG3-RPGV-2025-019), by the JC STEM Lab of Smart City funded by The Hong Kong Jockey Club Charities Trust under Contract 2023-0108, and by the Hong Kong SAR Government under the Global STEM Professorship and Research Talent Hub.}
\thanks{(Corresponding Author: Guang Hua and Yuguang Fang.)}
}



\maketitle

\begin{abstract}
Securing intellectual property (IP) in Machine Learning as a Service is critical yet challenging. While deep neural network watermarking serves as a standard defense against model extraction, existing Watermarking-as-a-Service paradigms face a triple challenge of \textit{intrusiveness}, \textit{privacy risks}, and \textit{inefficiency}. To address these challenges, we propose \textbf{N}on-intrusive \textbf{W}atermarking \textbf{a}s \textbf{a} \textbf{S}ervice (\textbf{NWaaS}), a holistic framework enabling trustworthy and efficient IP protection. We first introduce $\mathtt{ShadowMark}$, a novel watermarking algorithm that establishes a side-channel for ownership verification without modifying the model. It ensures zero performance degradation and eliminates the need for parameter-heavy fine-tuning as well as access to original training data, thereby addressing the intrusiveness and inefficiency inherent in existing approaches. Leveraging this non-intrusive property, we design a collaborative partitioning mechanism that allows model owners to offload self-defined partial layers, enabling a flexible trade-off between IP privacy and service cost. Furthermore, to mitigate latency from collaborative computing under high concurrency and enhance system resource utilization, we propose proportion disparity joint scheduling, a payload-balancing resource scheduling algorithm tailored to the heterogeneous constraints of edge-cloud environments. Extensive experiments demonstrate that NWaaS provides robust ownership verification across diverse continuous X-to-Image modalities, while ensuring secure owner privacy protection and superior system performance.
\end{abstract}

\begin{IEEEkeywords}
 Watermarking, Intellectual Property Protection, Privacy-Preserving Computing, Resource Scheduling.
\end{IEEEkeywords}

\begin{figure}[!t]
    \centering
    \includegraphics[width=\linewidth]{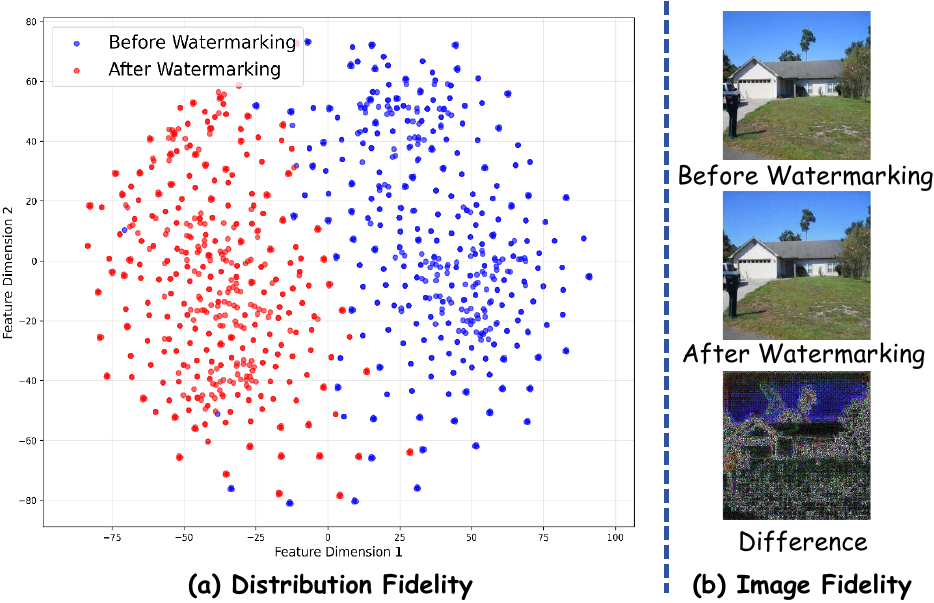}
    \caption{Visualization of fidelity control on feature level and image level for state-of-the-art box-free model watermarking \cite{Zhang2024Robust_Box_Free}, which aims to protect both the model and its output. Features are extracted using a pretrained VGG and visualized using t-SNE. The difference is taken between the images before and after watermarking, amplified 10$\times$ for better visibility.}
    \label{fig:fidelity_demonstration}
\end{figure}

\section{Introduction}
\IEEEPARstart{T}{he} mobile application market is witnessing a surge of intelligent apps, such as AI-powered photo editors and image generators, that rely on powerful, proprietary deep neural networks (DNNs) deployed on the cloud server as a Machine Learning as a Service (MLaaS). As the culmination of significant investment in big data, computational power, and human expertise, these cloud-based models represent core assets for developers, considered as valuable intellectual property (IP). However, this MLaaS paradigm is under serious threat from model extraction attack (also known as model stealing attack), where adversaries can systematically query the publicly accessible API to replicate a functionally similar surrogate model, significantly violating model owner's IP and resulting in model infringements and economic losses \cite{podhajski2026stealing, jin2026do, zhao2026trident}.

To counter this threat, DNN watermarking has emerged as the prevailing defense mechanism, enabling ownership verification by embedding owner's designated mark into the models. Since its inception \cite{Uchida2017Embedding_White_Box}, the field has attracted extensive research \cite{Li2021A_Survey} and even led to commercial Watermarking-as-a-Service (WaaS) platforms \cite{steg_ai, deepmark}, allowing model owners to protect IP without maintaining an in-house watermark embedding and extraction system. Despite these advancements, existing watermarking systems share three fundamental weaknesses that hinder their deployment in practical MLaaS environments.

\noindent\textbf{1) They are intrusive.} 
    Nearly all embedded watermarks rely on redundancies added to model parameters or architectures, inevitably resulting in performance degradation. For example, even state-of-the-art black-box watermarking methods still cause classification boundary shifts \cite{Hua2023Deep_Black_Box} and a similar degradation in both distribution, and image fidelity exists for the advanced box-free model watermarking as illustrated in Fig.~\ref{fig:fidelity_demonstration}, lowering the end-user's quality of experience (QoE) and affecting model evaluation. In addition, due to the inherent lack of interpretability in DNNs, such intrusiveness can induce unpredictable behavioral shifts, discouraging model owners from applying watermarks in modification-intolerant scenarios.
    
\noindent\textbf{2) They are unreliable and privacy-unfriendly.} 
    The current WaaS paradigm necessitates that model owners upload their complete, proprietary models to a third-party service provider, raising a significant trust issue that owners disclose their core IP (model architecture and parameters) to the provider. Furthermore, they lack reliable mechanisms to verify whether the returned model contains only the benign watermark or if malicious backdoors have been injected during the watermarking process.
    
\noindent\textbf{3) They are inefficient.} The watermarking process is resource-intensive, typically requiring the fine-tuning of the entire model. Crucially, to mitigate catastrophic forgetting during this process, access to the original training data is often mandatory. Transferring large-scale datasets and fine-tuning parameter-intensive models not only increases computational overhead but also introduces significant communication costs and data privacy risks.

To address these challenges, we propose Non-intrusive Watermarking as a Service (NWaaS), a holistic framework enabling trustworthy, privacy-preserving, and efficient IP protection. Our design philosophy follows a progressive logic.

\noindent\textbf{First, ensuring absolute fidelity and non-intrusiveness.} We propose $\mathtt{ShadowMark}$, a novel algorithm that establishes a side-channel for ownership verification without modifying the model to be protected. It augments the API with a lightweight, trainable key generator and decoder that extract the watermark from the unmodified model's outputs, achieving zero performance degradation and eliminating backdoor risks. Moreover, the training process only requires secret keys, random noise, and fixed out-of-distribution (OOD) data, completely preventing the need from accessing the original training data.

\noindent\textbf{Second, ensuring non-intrusiveness to privacy via collaborative partitioning.} Due to the non-intrusive nature of $\mathtt{ShadowMark}$, the watermarking process no longer requires modifying model parameters, so that the physical possession of the full model by the server is not mandatory. Building upon this, we design a collaborative partitioning mechanism that allows model owners to perform partial model submission. Owners can execute sensitive layers locally and only exchange intermediate values (activations and gradients) with the cloud server, allowing owners to flexibly control the trade-off between IP privacy and service cost by adjusting the partition.

\noindent\textbf{Third, enhancing service efficiency and reliability via adaptive scheduling.} While this collaborative architecture enhances privacy, the frequent exchange of intermediate data introduces complex dependencies between communication and computation. To ensure system efficiency and reliability under high concurrency, we integrate an adaptive resource allocation strategy, which has recently drawn attention in collaborative and split learning environments \cite{jeong2025collaborative, huang2025hyperjet}, and propose proportion disparity based joint scheduling (PDJS) algorithm, dynamically balancing communications, and server-side computing power, preventing congestion and ensuring low-latency service in diverse deployment scenarios.

Our contributions are summarized as follows.

\begin{enumerate}
\item We introduce $\mathtt{ShadowMark}$, a novel algorithm that overcomes the intrusive nature of existing methods. By establishing a key-driven side-channel, it achieves verifiable IP protection with zero performance degradation while eliminating the need of fine-tuning parameter-intensive models and accessing original training data.

\item We propose an NWaaS framework, a client-server system that resolves trust and privacy concerns. It features a collaborative partitioning mechanism that enables flexible trade-offs between IP privacy and service cost, eliminating the need of full model disclosure.

\item We develop an adaptive resource allocation strategy, implemented via our proposed PDJS algorithm. This mechanism optimizes the scheduling of heterogeneous resources, ensuring the system's efficiency (high system throughput and resource utilization) and reliability in real-world MLaaS environments.

\item We validate NWaaS through extensive experiments on continuous X-to-Image models across multiple modalities, demonstrating robust ownership verification under model extraction attacks, strong security against brute-force key ambiguity, and efficient resource scheduling via PDJS under realistic simulation conditions.
\end{enumerate}

\section{Related Work}
\subsection{Intrusive DNN Watermarking}
\label{sec:relatedwork_model_wm}
\begin{figure}[!t]
    \centering
    \includegraphics[width=\linewidth]{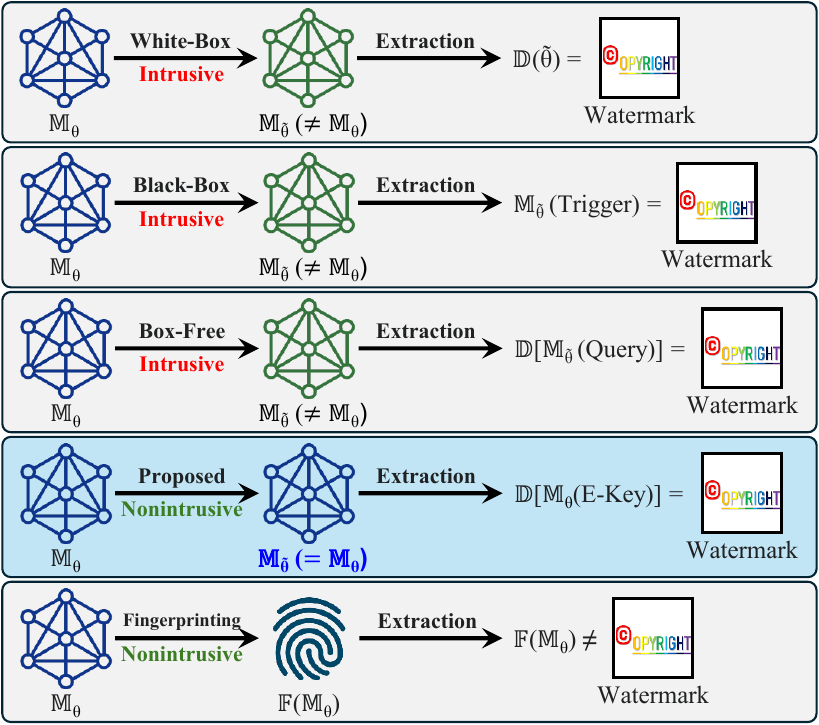}
    \caption{Conceptual comparison of the proposed non-intrusive DNN watermarking and related work, where $\mathbb{M}_\theta$ is the pretrained to-be-protected model, $\mathbb{M}_{\tilde{\theta}}$ is the protected model after watermarking, $\mathbb{F}$ is a fingerprinting model, $\mathbb{D}$ is the watermark decoder, and E-Key is encoded key. The Trigger, Query, and E-Key share the same dimension.}
    \label{fig:concept_compare}
\end{figure}

The existing DNN watermarking methods are intrusive in nature, which are summarized as follows. In addition, the difference between the proposed non-intrusive DNN watermarking and related work is demonstrated in Fig.~\ref{fig:concept_compare}.

\subsubsection{White-Box Watermarking} White-box methods directly modify the model's original layers \cite{Uchida2017Embedding_White_Box}, additional layers \cite{Fan2019Rethinking_White_Box,Cui2024Steganographic_White_Box}, special features \cite{Li2022Defending_White_Box, chen2025plugmark_White_Box}, parameters \cite{Tondi2024Robust_White_Box,Lv2023A_White_Box}, and etc. They are similar to the classic media watermarking, where the model parameters are analogous to the host media content. However, due to the requirement of the model’s internal content for watermark extraction, white-box methods have limited applicability to privacy-sensitive scenarios. 

\subsubsection{Black-Box Watermarking} Black-box methods \cite{liu2025attmark_Black_Box} achieve watermarking through the learning of specially designed trigger-response relations. The existing works have mainly focused on the black-box watermarking of deep classification models, and the trigger-label pairs can be out-of-distribution (OOD) samples with random labels \cite{Hua2024Unambiguous_Black_Box,Hua2023Deep_Black_Box,Adi2018Turning_Black_Box}, special-pattern-masked training samples with incorrect labels \cite{Zhang2018Protecting_Black_Box,Liu2024Robust_Black_Box,Jia2021Entangled_Black_Box}, or original samples with additional labels \cite{Li2020Protecting_Black_Box,Zhong2020Protecting_Black_Box}. Watermark is then extracted by observing the model’s responses to the predefined trigger set in a black-box manner.

\subsubsection{Box-Free Watermarking} Box-free methods \cite{Lin2024A_Box_Free, an2025decoder_Box_Free, an2025removing_Box_Free} embed the watermark into model outputs and use a dedicated decoder for watermark extraction. Watermarking can be established by jointly training the model and decoder \cite{Wu2021Watermarking_Box_Free,Lukas2023PTW_Box_Free} or fine-tuning the model with a pretrained and frozen decoder \cite{Lin2024A_Box_Free,Fei2024Wide_Box_Free}. Box-free methods are mainly applied to models generating high-entropic outputs, e.g., images. We also note that box-free watermarking can be fully decoupled from the model, which becomes DNN-based image watermarking.

More recently, Zheng \textit{et al.} \cite{nonintrusive_cyclegan} propose a non-intrusive watermarking method primarily for CycleGAN. Our work significantly extends the scope to a broader class of X-to-Image models and, more importantly, integrates it into a holistic \textit{Watermarking-as-a-Service} system (NWaaS) that addresses the privacy concerns and collaborative resource scheduling challenges, which have not been considered.

\subsubsection{Fingerprinting and Zero Watermarking}
There is a sizable body of work focusing on DNN fingerprinting with non-intrusive nature \cite{Zheng2022A_Fingerprinting,Hu2024VeriDIP_Fingerprinting,Cao2021IPGuard_Fingerprinting,Peng2022Fingerprinting_Fingerprinting,Lukas2021Deep_Fingerprinting,shao2025explanation_Fingerprinting}. The fundamental idea is to find a robust feature of the model, which can be the decision boundary \cite{Cao2021IPGuard_Fingerprinting}, adversarial attack sensitivity \cite{Peng2022Fingerprinting_Fingerprinting,Lukas2021Deep_Fingerprinting}, special layer parameters \cite{Zheng2022A_Fingerprinting}, or sensitivity to special training data \cite{Hu2024VeriDIP_Fingerprinting}. These features are directly governed by the inherent property of the protected model, which cannot protect model outputs or embed an owner-defined fingerprint-independent watermark. In contrast, NWaaS enables embedding owner-defined watermarks without relying on model-specific features. Another domain that links the feature or fingerprint to a watermark is the so-called zero watermarking \cite{Xia2021Color_Zero_Watermark}, whose essential component is the XOR scrambling between the feature and the owner-defined watermark. It then becomes an encryption scheme, deviating from the purpose of watermarking \cite{Cox2006Watermarking_Background}.

\subsection{Privacy-Preserving Computation in MLaaS}
In the field of MLaaS, privacy-preserving computation has attracted significant attention, primarily focusing on protecting data privacy rather than model IP. Federated Learning (FL) \cite{Tran2019Federated, Jiang2024Low_Parameter, wu2024fedbiot} enables distributed training by keeping raw data on local devices. However, it inherently requires deploying complete model replicas to clients and uploading model updates for aggregation, both of which pose severe threats of IP theft and are unsuitable for protecting proprietary models. To mitigate computational burdens while preserving data privacy, split learning (SL) \cite{lin2024efficient, Vepakomma2018Split} and its variants \cite{lin2025adaptsfl} partition the model into client-side and server-side segments, exchanging only intermediate activations and gradients. Although NWaaS shares the collaborative partitioning architecture with SL, its design philosophy is fundamentally different. SL is strictly data-driven, where the partition point is constrained to shallow layers to conceal raw inputs \cite{Vepakomma2018Split} or utilizes U-shaped structures to protect labels \cite{Yang2022Robust}. In contrast, our approach is IP-driven. By leveraging the non-intrusive nature of our watermarking, NWaaS decouples partitioning from data privacy constraints, allowing model owners to flexibly retain deep, IP-critical layers locally to prevent the server from reconstructing the functional model, whereas SL primarily aims to prevent the server from reconstructing data.

\section{System and Threat Models}
\label{sec:system_model}
We consider a real-world system where IP protection interacts with privacy concerns and resource constraints. This section details the operational entities, the collaborative architecture designed to satisfy privacy constraints, and the specific adversarial threats.
\subsection{System Model: A Collaborative Paradigm}
\label{subsec:system_model}
\begin{figure}[!t]
    \centering
    \includegraphics[width=1.0\linewidth]{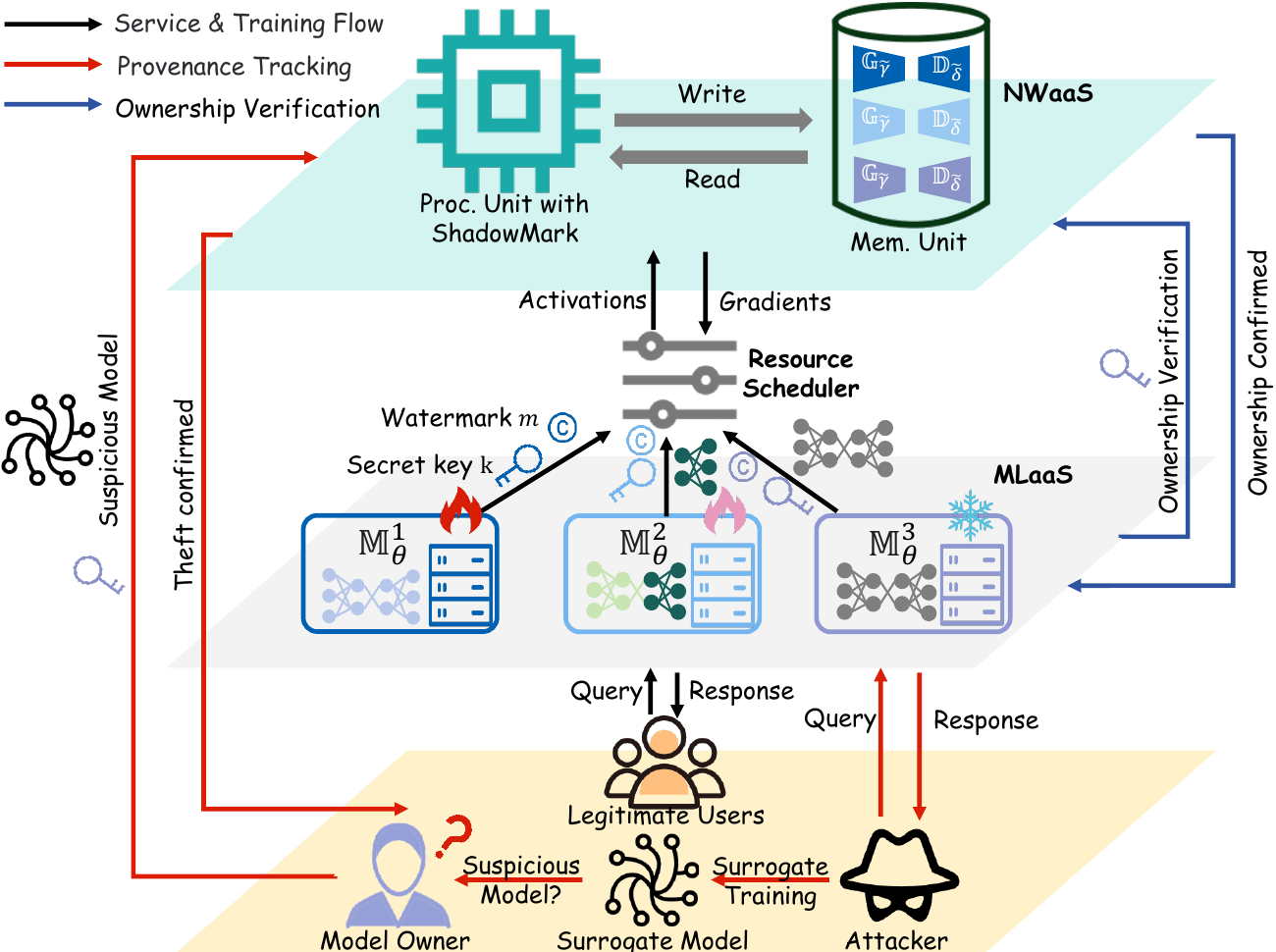}
    \caption{Overview of the integrated system comprising NWaaS, MLaaS, and end-users. Black lines denote the standard machine learning service and collaborative training flows, blue lines represent the ownership verification process for deployed authorized models, and red lines illustrate the provenance tracking against suspicious surrogate models.}
    \label{fig:architecture}
\end{figure}

As illustrated in Fig.~\ref{fig:architecture}, the NWaaS system operates as a dynamic, collaborative process involving three key parties.

\subsubsection{The Model Owner ($o_n$)}
The system centers around the model owner, denoted as $o_n$ ($n \in \{1, \dots, N\}$), who possesses a high-value proprietary DNN, $\mathbb{M}_{\theta}$, consisting of $L_n$ layers. This model represents the owner's core IP. The owner aims to protect $\mathbb{M}_{\theta}$ via watermarking but typically operates on edge infrastructure with limited computational power. To verify ownership, the owner generates a secret key $\mathsf{k} \in \mathcal{K}$ which is kept private from external adversaries.
Crucially, the owner faces a dilemma: they require the provider's computing power for efficient processing but are unwilling to disclose the specific sensitive parameters or architecture of $\mathbb{M}_{\theta}$.
To resolve this, $o_n$ participates in a collaborative training protocol to train the watermarking modules ($\mathbb{G}_{\gamma}, \mathbb{D}_{\delta}$) without surrendering the full control of the model.

\subsubsection{The NWaaS Provider (Server)}
The server offers a managed watermarking service powered by GPU clusters. It maintains the trainable components ($\mathbb{G}_{\gamma}, \mathbb{D}_{\delta}$) and conducts the collaborative training. It schedules the resource allocation for $N$ model owners but remains oblivious to the internal parameters of the model portions retained by the owners. The server is assumed to be honest-but-curious, faithfully executing the protocol while potentially attempting to infer the model's IP.

\subsubsection{The End-Users}
The deployed model is accessible via an API to legitimate end-users, who submit queries to receive inference results. While legitimate users seek normal services, this public access channel may also be exploited by potential adversaries to launch model extraction attacks.

\subsubsection{Flexible Partitioning Mechanism}
\label{subsubsec:partitioning}
To strike a privacy-efficiency balance between the owner and the server, we introduce an IP-driven flexible partitioning mechanism, which fundamentally differs from the traditional SL.

\noindent\textbf{IP vs. Data Privacy:} In SL, the partition is constrained by data privacy, mandating that input layers remain local to hide raw training data. In contrast, our NWaaS utilizes $\mathtt{ShadowMark}$ which relies on secret key, random noise, and OOD data rather than private training data. Since the verification inputs are non-sensitive to the raw training data, the strict constraint of keeping input layers local is eliminated.

\noindent\textbf{Arbitrary Cut Point ($l_n$):} The owner  $o_n$ has complete freedom to define the trust boundary. They can select a cut point $l_n$ to retain any portion of the model locally based on their IP valuation.

\noindent\textbf{Full Flexibility,} allowing for diverse configurations:
    \begin{enumerate}
        \item \textit{Shallow-Offload:} Offloading non-sensitive shallow layers to the server while keeping deep, IP-critical layers local.
        \item \textit{Deep-Offload:} Offloading deep layers if the unique feature extraction in shallow layers is the core IP.
        \item \textit{Full-Local Execution:} In extreme cases where the entire architecture is highly sensitive, $o_n$ can execute the full model locally ($l_n = L_n$), using the server solely for coordinating the watermarking gradients, thereby achieving maximum privacy.
    \end{enumerate}

\begin{figure}[!t]
    \centering
    \includegraphics[width=.994\linewidth]{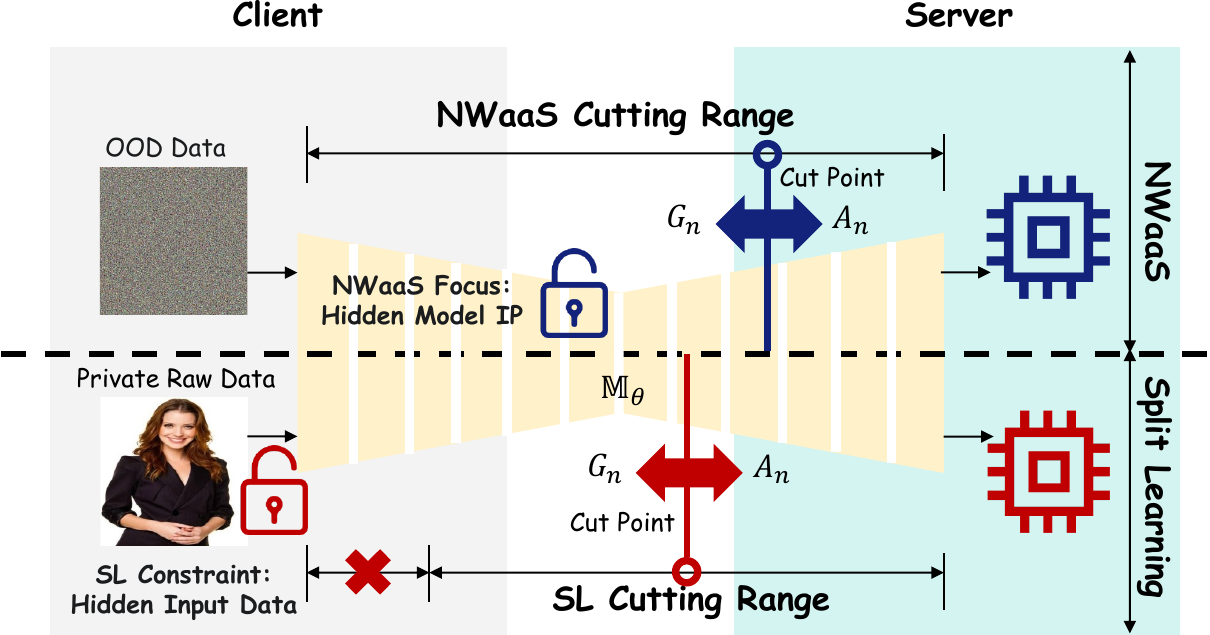}
    \caption{Illustration of our proposed collaborative partitioning mechanism contrasting NWaaS with classic split learning (SL).}
    \label{fig:compare_sl}
\end{figure}

\noindent Without loss of generality, this paper focuses on the deep-offload configuration. The distinction between our proposed collaborative partitioning mechanism and classic SL in this scenario is illustrated in Fig.~\ref{fig:compare_sl}. During this collaboration, only the intermediate activations $A_n$ and backward gradients $G_n$ at the cut point are exchanged. The server bears the computational load $\eta_n$ for the offloaded parts, balancing privacy and  the computational cost.

\subsection{Threat Model: A War on Two Fronts}
Our security analysis operates under a comprehensive threat landscape specifically targeting X-to-Image modalities. Before detailing the specific threats, we define the malicious user (adversary). This entity possesses access rights similar to a normal end-user and interacts with the deployed API. However, unlike legitimate users, their purpose is not service consumption but model extraction.

\subsubsection{External Threat (Model Extraction)}
We make the following assumptions about the adversary aiming to steal the functionality of $\mathbb{M}_{\theta}$.
\begin{itemize}
    \item \textbf{Capability:} We assume an adversary-permissive threat model to demonstrate the effectiveness and robustness of our proposed methods, where the attacker has black-box access to the deployed API, with unlimited capability to query $\mathbb{M}_{\theta}$ to train a surrogate model $\mathbb{S}_{\vartheta}$ (Watermark removal attack). Crucially, this black-box access is strictly limited to the target model $\mathbb{M}_{\theta}$. The attacker cannot access the key generator $\mathbb{G}_{\gamma}$, the decoder $\mathbb{D}_{\delta}$, or any internal verification feedback, preventing them from executing black-box search optimizations.
    \item \textbf{Goal:} To obtain a functional copy of the model while invalidating the ownership verification (i.e., making the watermark $\hat{m}$ undetectable with the owner's key $\mathsf{k}$).
    \item \textbf{Knowledge:}  Following Kerckhoffs's principle, the watermarking algorithm and the statistical distribution of the key space (e.g., sampling from a standard normal distribution $\mathcal{N}(0, I)$) are publicly known, but the specific instance of the key $\mathsf{k}$ is kept strictly private.
\end{itemize}

\subsubsection{Internal Threat (The Honest-but-Curious Server)}
\label{sec:internal_threat}
We model the NWaaS Server as \textit{Honest-but-Curious} (HBC). This assumption is critical to justify our flexible partitioning design:
\begin{itemize}
    \item \textbf{Honest:} The server faithfully executes the protocols ($\mathtt{ShadowMark}$ training, PDJS scheduling) and returns correct calculation results. Note that due to the non-intrusive property where sensitive model parameters are kept frozen, the server is technically incapable of injecting backdoors into $\mathbb{M}_{\theta}$, resolving the trust issue inherent in the traditional WaaS.
    
    \item \textbf{Curious:} The server may be interested in retaining the uploaded model copies to conduct secondary utilization (e.g., internal analytics, service optimization, or knowledge accumulation, but surrogate training is not included) for its own benefit. While these activities may not directly affect the owner's service, they constitute unauthorized usage of proprietary assets.
\end{itemize}
To preserve IP exclusivity, the model owner refuses to trust the provider with the full model storage. Our flexible partitioning ensures that a complete, functional copy of $\mathbb{M}_{\theta}$ never physically exists on the server side, thereby technically precluding any possibility of unauthorized retention.

\subsection{Design Goals}
\begin{enumerate}
    \item \textbf{Non-intrusiveness (Fidelity):} The protection must be transparent, keeping original parameters $\theta$ unchanged to ensure zero performance degradation.
    \item \textbf{Verifiable Ownership:} The system must robustly verify ownership using the trained $\left(\mathbb{G}_{\tilde{\gamma}}, \mathbb{D}_{\tilde{\delta}}\right)$ and a secret key $\mathsf{k}$, against surrogate models.
    \item \textbf{Owner-Defined Privacy:} The system must support flexible partitioning, allowing owners to hide any specific layers (not limited to input layers) from the server based on their specific sensitivity requirements.
    \item \textbf{System Efficiency:} Given the communication overhead of collaborative computing, the system must employ adaptive scheduling (PDJS) to prevent congestion, enhance overall resource utilization, and maximize system throughput (tasks processed per unit time) under high concurrency.
\end{enumerate}

\section{NWaaS: Algorithm and System Design}
This section details the architectural design of our NWaaS. We first provide a brief and unified review of existing model watermarking definitions and then introduce the core algorithm, $\mathtt{ShadowMark}$, which establishes a verifiable side-channel for IP protection. Subsequently, we present the collaborative partitioning mechanism designed to balance privacy and computational cost. Finally, we present the PDJS algorithm, an adaptive resource allocation strategy tailored to resolve resource contention and ensure system efficiency under high concurrency.
\subsection{Preliminary: Model Watermarking}
\label{sec:pre_model_wm}
As discussed in Section \ref{sec:relatedwork_model_wm}, current model watermarking approaches necessitate directly (white-box) or indirectly (black-box and box-free) modifications to the model parameters or architectures to facilitate watermark embedding. Existing algorithms define a pair of functions, $\mathtt{Encode}(\cdot)$ and $\mathtt{Decode}(\cdot)$, to embed and extract an owner-defined watermark $m \in \mathcal{M}$ (where $\mathcal{M}$ denotes the watermark space, e.g., $\mathcal{M} = [0,1)^{3\times H\times W}$ for images), utilizing optional secret keys $\mathsf{k_e}$ and $\mathsf{k_d}$. 
For simplicity, we focus on model parameter modification in this context. With the above settings, we have:

\noindent $\bullet$ White-box Model Watermarking:

1. Embed: $\mathbb{M}_{\tilde{\theta}} \leftarrow \mathtt{Encode}(\mathbb{M}_{\theta}, m, \mathsf{k_e})$,

2. Extract: $\hat{m} \leftarrow \mathtt{Decode}(\mathbb{D}, \mathbb{M}_{\tilde{\theta}},\mathsf{k_d})$,\\
\noindent where $\mathbb{D}$ is the watermark decoder;

\noindent $\bullet$ Black-box Model Watermarking:

1. Generate Pair: $(\mathsf{t}, \mathsf{r}) \leftarrow \mathtt{TriggerGen}(m)$,

2. Embed: $\mathbb{M}_{\tilde{\theta}} \leftarrow \mathtt{Encode}\left(\mathbb{M}_{\theta}, (\mathsf{t}, \mathsf{r}), \mathsf{k_e}\right)$,

3. Extract: $\hat{\mathsf{r}} \leftarrow \mathtt{Decode}(\mathbb{M}_{\tilde{\theta}}, \mathsf{t}, \mathsf{k_d})$,

4. Map: $\hat{m} \leftarrow \mathtt{Map}\left(\mathsf{t}, \hat{\mathsf{r}}\right)$,\\
\noindent where $(\mathsf{t}, \mathsf{r})$ is the owner-defined backdoor trigger-response pair, also known as the watermarking key, $\mathtt{TriggerGen}(\cdot)$ is a public one-to-many function that maps the watermark $m$ to the watermarking key, and $\mathtt{Map}(\cdot)$ is a public many-to-one function that maps the key back to the watermark $m$;

\noindent $\bullet$ Box-free Model Watermarking:

1. Embed: $\mathbb{M}_{\tilde{\theta}} \leftarrow \mathtt{Encode}(\mathbb{M}_{\theta}, x, m, \mathbb{D}, \mathsf{k_e})$,

2. Extract: $\hat{m} \leftarrow \mathtt{Decode}(\mathbb{D}, \mathbb{M}_{\tilde{\theta}}(x), \mathsf{k_d})$,\\
\noindent where $x\in\mathcal{X}$ is the query from training data distribution. 

\noindent\textbf{Remark on Intrusiveness:} A critical observation is that all aforementioned paradigms involve the transition from the original model $\mathbb{M}_{\theta}$ to a modified version $\mathbb{M}_{\tilde{\theta}}$. This modification inevitably alters the model's parameters, introducing the risk of performance degradation and fidelity loss, which we refer to as the intrusive nature of existing watermarking.

For watermark verification, the extracted output $\hat{m}$ is compared with the original watermark $m$, given by
\begin{align}
1 \leftarrow \mathtt{Verify}(\hat{m}, m) & \text{ if } \mathtt{Criterion}(\hat{m}, m) > \eta, \nonumber \\
& \text{ and } 0 \text{ otherwise}, \label{eq:verify}
\end{align}
where $\eta$ is the decision threshold related to the value range of $\mathtt{Criterion}$. Existing works have incorporated different $\mathtt{Criterion}$ functions. For binary $m$ such as a binary sequence or a black-white watermark, bit error rate (BER) or equivalently bit accuracy (BAC) has been used, and the corresponding threshold $\eta$ can be, e.g., $\mathtt{BAC} > 0.85$ \cite{fernandez2023stable_Box_Free}, $\mathtt{BAC} > 0.65$ \cite{Lukas2023PTW_Box_Free}, $\mathtt{BAC} > 0.63$ \cite{Zhao2024Invisible_Removal_Attack}, $\mathtt{BAC} > 0.69$ for a $200$-bit watermark \cite{Lin2024A_Box_Free}, and $\mathtt{BAC} > 0.61$ for a $100$-bit watermark \cite{Fei2024Wide_Box_Free}. For image watermarks, peak signal-to-noise ratio (PSNR) or normalized cross-correlation (NCC) has been used, e.g., $\mathtt{PSNR}>35$ dB \cite{Zhang2024Suppressing_Box_Free}, $\mathtt{NCC}>0.95$ \cite{Zhang2022Deep_Box_Free,Zhang2024Robust_Box_Free,Chen2024High_Box_Free}. In addition, we note that in \cite{Wu2021Watermarking_Box_Free}, both BAC and PSNR are used without an $\eta$, while in \cite{Baluja2020Hiding}, the verification is based on database retrieval.

\subsection{Core Algorithm: $\mathtt{ShadowMark}$}
\label{sec:core_algorithm}

We now introduce the details of our $\mathtt{ShadowMark}$.

\subsubsection{Main Idea}
 Our fundamental hypothesis is that for a high-capacity DNN $\mathbb{M}_\theta$, the watermark extraction mapping can be constructed without altering the model parameters $\theta$. 

Intuitively, this seems paradoxical: \textbf{\emph{how can ownership be verified if the watermark is never embedded?}} 
We resolve this by shifting the paradigm from watermark embedding to manifold discovery.
Recall that standard box-free watermarking trains a decoder $\mathbb{D}_\delta$ to map the output of a modified model $\mathbb{M}_{\tilde{\theta}}$ to $m$. 
Our insight is that $\mathbb{M}_{\theta}$, as a highly non-linear function, naturally possesses a complex manifold capable of mapping diverse inputs to diverse regions in the output space. Instead of modifying $\mathbb{M}_{\theta}$ to fit a specific input-output pair (which causes intrusiveness), we propose to discover a latent trigger-key pair in the input space that naturally maps to the target watermark $m$ through the frozen $\mathbb{M}_{\theta}$ and a learnable decoder.

\begin{figure}[!t]
    \centering
    \includegraphics[width=.994\linewidth]{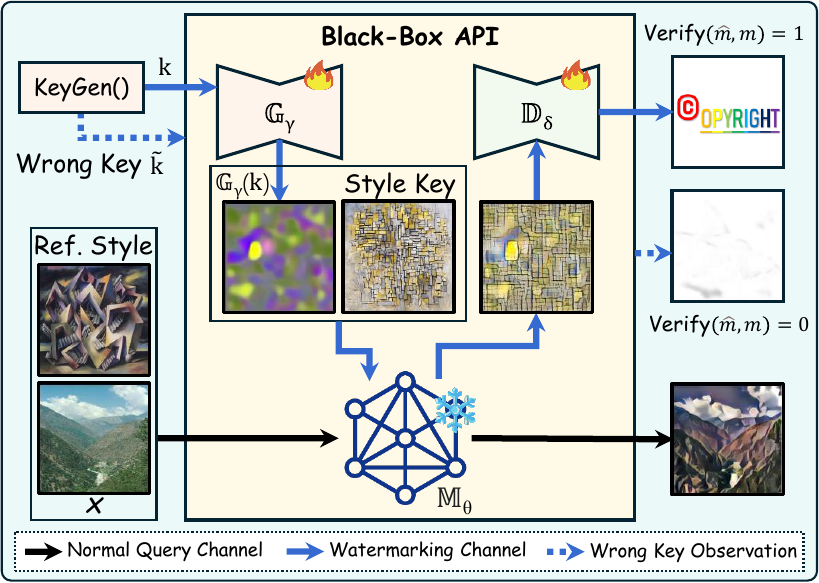}
    \caption{Flowchart of the proposed $\mathtt{ShadowMark}$ using protecting image style transfer model as an example. 
    For the ease of visualization, the intermediate results in the black-box API from a wrong key are omitted. The Ref. Style image is model-dependent additional input, while the Style Key is also model-dependent and an additional fixed secret key.}
    \label{fig:shadowmarkFlowchart}
\end{figure}

This process can be formalized as finding a specific trajectory through the fixed model manifold. However, directly training a decoder $\mathbb{D}_\delta$ on natural inputs leads to a trivial mapping where $\mathbb{D}_\delta$ simply memorizes $m$ regardless of the input (a phenomenon often resembling overfitting in standard training). 
To overcome this and establish a rigorous cryptographic binding, we introduce a key-driven mechanism. As illustrated in Fig.~\ref{fig:shadowmarkFlowchart},  where image style transfer model is used as an example, $\mathtt{ShadowMark}$ augments the inference pipeline with two distinct pathways: the normal inference channel (black arrows) and the watermarking channel (blue arrows).
Specifically, we employ a trainable key generator, $\mathbb{G}_\gamma$, which transforms an owner-defined secret key $\mathsf{k}$ into a specific trigger query. The system effectively learns a composite function $\mathbb{D}_\delta(\mathbb{M}_\theta(\mathbb{G}_\gamma(\cdot)))$ such that it acts as an identity mapping from $\mathsf{k}$ to $m$, while keeping $\mathbb{M}_\theta$ frozen.

This design achieves non-intrusiveness, as $\mathbb{M}_\theta$ remains strictly untouched. The formal definition of the $\mathtt{ShadowMark}$ is as follows:

\noindent $\bullet$ $\mathtt{ShadowMark}()$:

1. Generate Secret Key: $\mathsf{k} \leftarrow \mathtt{KeyGen()}$,

2. Embed: $\left(\mathbb{G}_{\tilde{\gamma}}, \mathbb{D}_{\tilde{\delta}}\right) \leftarrow \mathtt{Encode}\left(\mathbb{G}_{{\gamma}}, \mathbb{D}_{{\delta}}, \mathbb{M}_\theta, \mathsf{k}, m\right)$, 

3. Extract: {\small{$\hat{m} \leftarrow \mathtt{Decode}(\mathbb{G}_{\tilde{\gamma}}, \mathbb{D}_{\tilde{\delta}}, \mathbb{M}_{{\theta}}, \mathsf{k})$}},\\
By decoupling the watermarking logic (handled by $\mathbb{G}_{\tilde{\gamma}}$ and $\mathbb{D}_{\tilde{\delta}}$) from the model ($\mathbb{M}_\theta$), $\mathtt{ShadowMark}$ establishes a robust, verifiable side-channel. The API serves as a gatekeeper, determining whether to execute normal inference on natural queries $x$ or watermark verification on $\mathsf{k}$, ensuring that the watermarking process remains transparent to standard end-users while providing strong ownership proof.

\subsubsection{Implementation}
To implement $\mathtt{ShadowMark}$, given an owner-defined secret key $\mathsf{k}\in \mathcal{K}$ and watermark $m \in \mathcal{M}$, it then should follow that
\begin{align}
    & \mathtt{Verify}(\mathtt{ShadowMark}.\mathtt{Decode}(\mathsf{k}), m) = 1, \label{eq:true}\\
    & \mathtt{Verify}(\mathtt{ShadowMark}.\mathtt{Decode}(\tilde{\mathsf{k}}), m)= 0, \forall \tilde{\mathsf{k}} \neq \mathsf{k}. \label{eq:false}
\end{align}
 We note that simultaneously establishing both equalities is essential to control false and miss verifications. To achieve this, our loss functions for training $\mathtt{ShadowMark}$ are designed as follows.

\textbf{(1) Loss for Correct Key $\mathsf{k} \in \mathcal{K}$.} To achieve (\ref{eq:true}), we use the mean squared error (MSE) loss
\begin{equation}
 \label{eq:loss_k}
    \mathcal{L}_{\mathsf{k}} = \|\mathbb{D}_{\delta}\left(\mathbb{M}_{{\theta}}\left( \mathbb{G}_{\gamma}(\mathsf{k}) \right)\right) - m \|_2^2,
\end{equation}
to ensure that $\mathsf{k}$ can be successfully mapped to $m$. This is the main loss function to establish the watermarking channel.

\textbf{(2) Loss for Wrong Key $\tilde{\mathsf{k}} \neq \mathsf{k}$.} In this situation, our objective is to maximize the Euclidean distance between the decoder output and $m$, so as to achieve (\ref{eq:false}), for which we propose the following loss function
\begin{equation}
    \label{eq:loss_wrong_k}
    \mathcal{L}_{\tilde{\mathsf{k}}} = \frac{1}{ \|\mathbb{D}_{\delta}\left(\mathbb{M}_{{\theta}}\left(  \mathbb{G}_{\gamma}(\tilde{\mathsf{k}}) \right)\right) - m \|_2^2+\epsilon},
\end{equation}
where $0 < \epsilon \ll 1$ is a small constant for numerical stability. It is worth noting that the intuitive loss $-\|\mathbb{D}_{\delta}\left(\mathbb{M}_{{\theta}}\left( \mathbb{G}_{\gamma}(\tilde{\mathsf{k}}) \right)\right) - m \|_2^2$ is not suitable as the learning will be led to minimizing it towards $- \infty$, making $\mathcal{L}_{\tilde{\mathsf{k}}}$ ineffective.

\textbf{(3) Loss for Refining $\mathbb{D}_\delta$.} 
A robust watermarking system must not hallucinate watermarks on normal end-user inputs. We treat natural queries $x \in \mathcal{X}$ as negative samples. Since $\mathbb{M}_\theta$ is frozen, we train the decoder $\mathbb{D}_\delta$ to differentiate between the watermark-carrying latent code (from $\mathbb{G}_\gamma(\mathsf{k})$) and natural feature maps (from $x$). The loss is defined as:
\begin{equation}
    \label{eq:loss_ood}
    \mathcal{L}_x = \frac{1}{\|\mathbb{D}_{\delta}\left(\mathbb{M}_{{\theta}}\left(x \right)\right) - m \|_2^2 + \epsilon},
\end{equation}
which follows that $0 \leq \mathcal{L}_x \leq 1/\epsilon$. We note that the encoded key $\mathbb{G}_{\gamma}(\mathsf{k})$ is a noise-like or abstract (see Fig.~\ref{fig:shadowmarkFlowchart} for example) image, for which the normal query $x$ can be considered as an OOD sample ($x$ does not belong to the original training dataset of $\mathbb{M}_{\theta}$). Therefore, $\mathcal{L}_x$ refines the discriminative capability of $\mathbb{D}_\delta$, effectively separating the watermark manifold from the natural inference manifold.

\textbf{Overall Loss.} With the above settings, the overall learning process for $\mathtt{ShadowMark}$ is given by
\begin{align}
    \left(\mathbb{G}_{\tilde{\gamma}}, \mathbb{D}_{\tilde{\delta}}\right) = & \mathtt{ShadowMark}.\mathtt{Encode}\left(\mathbb{G}_{{\gamma}}, \mathbb{D}_{{\delta}}, \mathbb{M}_\theta, \mathsf{k}, \tilde{\mathsf{k}}, x, m\right) \notag\\
    = &  \arg  \mathop {\min }\limits_{\mathbb{G}_\gamma, \mathbb{D}_\delta}  \mathcal{L}_{\mathsf{k}} + \mathcal{L}_{\tilde{\mathsf{k}}} + \mathcal{L}_x,
\end{align}
where the weighting parameters are omitted since all loss functions share the same scale. We note that since $\mathsf{k}$ is owner-defined, $\tilde{\mathsf{k}}$ is randomly sampled noise, and OOD sample $x$ is reusable for all watermarking embedding once collected by the NWaaS server. Therefore, $\mathtt{ShadowMark}$ is original-training-data-free, significantly enhancing both the model owner’s data privacy and the efficiency of data utility.

\subsubsection{Verification Process}
\label{sec:verification}

As established in the design goals, the model owner requires a robust mechanism to verify ownership without compromising model integrity. Leveraging the non-intrusive nature of $\mathtt{ShadowMark}$, the verification process inherently guarantees that the deployed model $\mathbb{M}_{\theta}$ remains mathematically equivalent to the original clean model, thereby eliminating risks of backdoor injection. To operationalize this in a trustless MLaaS environment, the verification protocol is designed with three strict principles:
1) \textit{Standardized Protocol}: The verification follows a public, deterministic procedure (as shown in Fig.~\ref{fig:shadowmarkFlowchart}, ensuring that no entity can manipulate the verification logic.
2) \textit{Channel Decoupling}: Normal end-user's inference queries and the owner's verification queries operate on decoupled channels within the cloud server, preventing service interference.
3) \textit{Key Sovereignty}: Following Kerckhoffs's Principle, while the watermark extraction algorithm is public, the specific trigger generation relies solely on a private secret key $\mathsf{k} \in \mathcal{K}$. This key is generated and exclusively held by the model owner, serving as the root of trust for claiming ownership over both the original deployment and any potential surrogate models.

However, a critical challenge arises in verifying stolen models (surrogates). As discussed in Section~\ref{sec:pre_model_wm}, existing criteria vary significantly in stringency (e.g., $\mathtt{NCC}>0.95$ \cite{Zhang2022Deep_Box_Free} vs. $\mathtt{BAC}>0.61$ \cite{Fei2024Wide_Box_Free}). We observe that surrogate models $\mathbb{S}_{\tilde{\vartheta}}$ obtained via extraction attacks often suffer from feature degradation and domain shifts. Consequently, applying a stringent absolute threshold (like $0.95$) to a degraded surrogate may lead to false negatives.

To address this, we propose a \textit{dual-mode verification strategy} that distinguishes between integrity verification (for the authorized model) and provenance tracking (for suspicious surrogates).

\textbf{(1) Ownership Verification for $\mathbb{M}_{\theta}$:}
For the deployed model where fidelity is preserved, we adhere to the stringent standard to ensure high confidence:
\begin{equation}
    \mathtt{Verify}_{\text{orig}}(\hat{m}, m) = 
    \begin{cases} 
    1, & \text{if } \mathtt{NCC}(\hat{m}, m) > 0.95, \\
    0, & \text{otherwise}.
    \end{cases}
    \label{eq:verify_m}
\end{equation}

\textbf{(2) Provenance Tracking for $\mathbb{S}_{\tilde{\vartheta}}$:}
For verifying suspicious surrogates, we exploit the key-sensitivity of $\mathtt{ShadowMark}$. We hypothesize that even a degraded surrogate model retains the learned manifold structure of the watermark trigger, reacting significantly stronger to the correct key $\mathsf{k}$ than to a random noise key $\check{\mathsf{k}}$. We propose the normalized cross-correlation difference (NCCD) to cancel out the baseline noise and measure this relative sensitivity:
\begin{equation}\label{eq:nccd}
    \mathtt{NCCD}(\hat{m}, \check{m}, m) = \mathtt{NCC}(\hat{m},m) - \mathtt{NCC}(\check{m},m),
\end{equation}
where $\hat{m}$ is extracted using the owner's key $\mathsf{k}$, and $\check{m} = \mathbb{D}_{\tilde{\delta}}(\mathbb{S}_{\tilde{\vartheta}}(\mathbb{G}_{\tilde{\gamma}}(\check{\mathsf{k}})))$ denotes the output from a randomly sampled invalid key $\check{\mathsf{k}} \neq \mathsf{k}$. By measuring the differential response, NCCD provides a robust metric invariant to the absolute performance degradation of the surrogate. The verification criterion is defined as:
\begin{equation}
    \mathtt{Verify}_{\text{surr}}(\hat{m}, \check{m}, m) = 
    \begin{cases} 
    1, & \text{if } \mathtt{NCCD}(\hat{m}, \check{m}, m) > 0.5, \\
    0, & \text{otherwise},
    \end{cases}
    \label{eq:verify_s}
\end{equation}
where the threshold $0.5$ here is an empirical value. This indicates that the correlation with the correct watermark must significantly exceed the baseline correlation of a random query.

\subsection{Collaborative Partitioning Mechanism}
\label{sec:partitioning}

In the NWaaS system, the server concurrently serves a massive number of model owners, denoted as $o_n$ ($n=1, \dots, N$). Each owner operates under heterogeneous device constraints and distinct IP privacy requirements. To enable efficient watermarking without compromising the confidentiality of the full model architecture and parameters, we introduce a collaborative partitioning mechanism. This mechanism transforms the rigid watermarking task into a tunable workload, determining the specific computation and communication profile that the central scheduler must manage.

\subsubsection{Flexible Partial Model Submission}
Let owner $o_n$ possess a proprietary model $\mathbb{M}_{\theta}^{(n)}$ consisting of $L_n$ layers with total computational complexity $C_n$ (measured in FLOPs). Unlike traditional approaches that necessitate full model uploading, NWaaS allows $o_n$ to select a strategic cut point $l_n \in \{0, \dots, L_n\}$ to partition $\mathbb{M}_{\theta}^{(n)}$ into two segments:
\begin{itemize}
    \item \textbf{Local Segment ($\mathbb{M}_{\text{loc}}^{(n)}$):} Comprising layers $1$ to $l_n$, which is retained and executed on the owner's local trusted device.
    \item \textbf{Remote Segment ($\mathbb{M}_{\text{rem}}^{(n)}$):} Comprising layers $l_n+1$ to $L_n$, which is offloaded to the NWaaS server for collaborative processing.
\end{itemize}
Consistent with the system model described in Section~\ref{subsubsec:partitioning}, without loss of generality, we focus on the \textbf{Deep-Offload} strategy in this paper. In this configuration, the specific parameters and architecture of $\mathbb{M}_{\text{loc}}^{(n)}$ remain physically isolated, and only the intermediate activations $A_n$ (forward pass) and backward gradients $G_n$ (backward pass) at the cut point are exchanged with the server.

\subsubsection{Owner-Side Workload Determination}
The selection of the cut point $l_n$ directly dictates the distribution of resources, serving as the input for the subsequent global scheduling. We formulate the dependency of privacy and system cost on $l_n$ as follows.

\textbf{Privacy Metric ($\mathcal{P}_n$):}
For owner $o_n$, the privacy metric $\mathcal{P}_n(l_n)$ represents the ratio of sensitive IP (e.g., architectural depth or parameter count) kept locally. A higher $l_n$ indicates that a larger portion of the model $\mathbb{M}_{\theta}^{(n)}$ is hidden from the server, satisfying a strictly higher privacy standard.

\textbf{Cost Profile ($\eta_n, A_n, G_n$):}
For a chosen $l_n$, the resulting workload is characterized by.
\begin{itemize}
    \item \textbf{Communication Load:} The size of data exchanged depends on the output tensor dimensions at layer $l_n$, denoted as $S_n(l_n)$. Thus, the upload size $A_n(l_n)$ and download size $G_n(l_n)$ are determined based on $S_n(l_n)$.
    \item \textbf{Computation Split:} Let $\Phi_n(l)$ denote the cumulative FLOPs of $\mathbb{M}_{\theta}^{(n)}$ up to layer $l$. The computational load ratio offloaded to the server, denoted by $\eta_n$, is formulated as:
    \begin{equation}
        \eta_n(l_n) = 1 - \frac{\Phi_n(l_n)}{C_n}.
    \end{equation}
\end{itemize}

\textbf{Optimal Cut Point Selection:}
Before entering the scheduling queue, each owner $o_n$ determines their optimal cut point $l_n^*$ locally. The goal is to minimize the estimated processing latency while strictly satisfying a personalized IP privacy constraint $\mathcal{P}_{\text{req}}^{(n)}$. This is formulated as:
\begin{equation}
\label{eq:local_opt}
    l_n^* = \arg \min_{l \in \{0, \dots, L_n\}} \left( \frac{\Phi_n(l)}{\omega_n} + \frac{C_n - \Phi_n(l)}{\omega_s} + \frac{2 S_n(l)}{B_{\text{est}}} \right),
\end{equation}
subject to $\mathcal{P}_n(l) \geq \mathcal{P}_{\text{req}}^{(n)}$, where $\omega_n$ and $\omega_s$ represent the computational capacities of the owner local device and the NWaaS server, respectively, and $B_{\text{est}}$ is the estimated network bandwidth. The factor $2$ accounts for both uplink ($A_n$) and downlink ($G_n$) transmission.

Once $l_n^*$ is determined, the specific workload profile $\{A_n, G_n, \eta_n\}$ is fixed and submitted to the NWaaS server. These aggregated profiles from $N$ owners form the heterogeneous inputs for the PDJS algorithm detailed in the next section.

\begin{figure}[!t]
    \centering
    \includegraphics[width=1.0\linewidth]{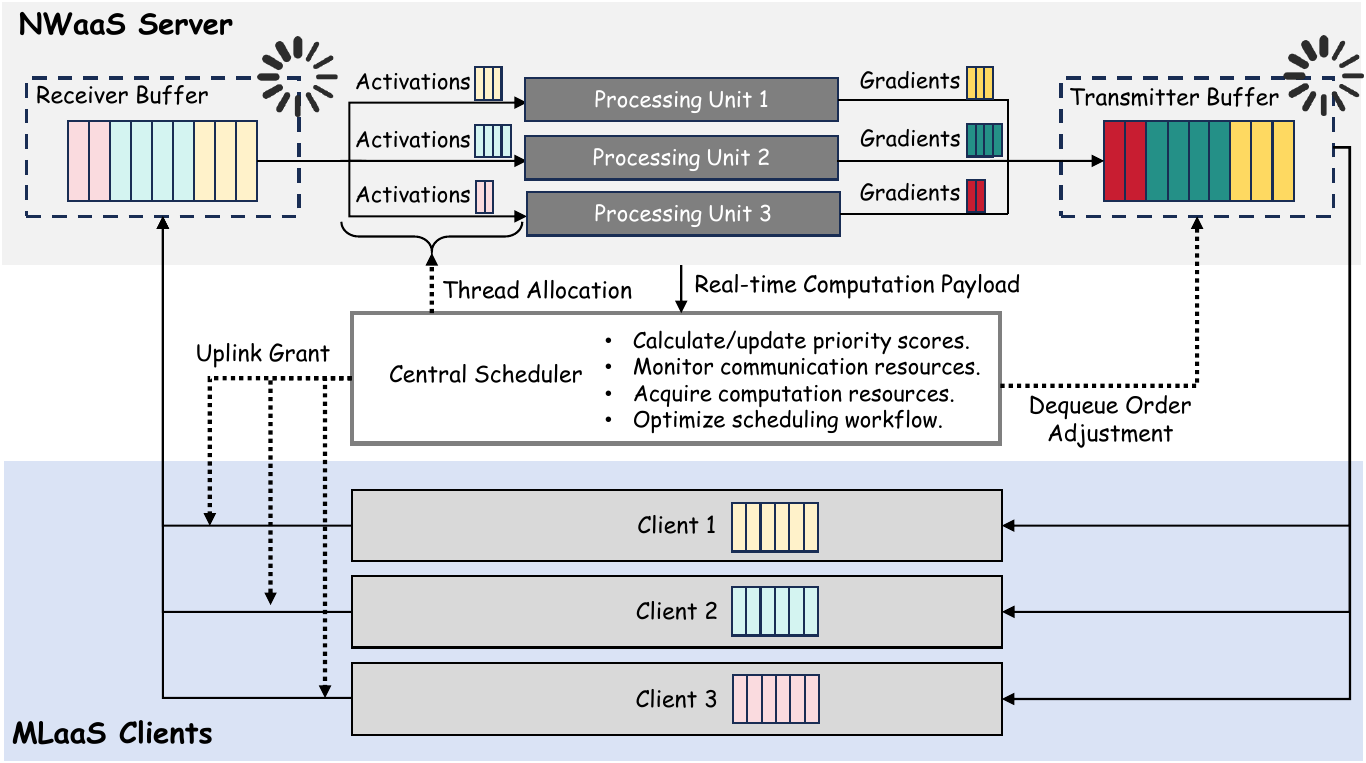}
    \caption{Illustration of the computation and communication model of NWaaS, where the client corresponds to the model owner $o_n$.}
    \label{fig:scheduler}
\end{figure}

\subsection{Proportion Disparity Joint Scheduling (PDJS)}
\label{sec:pdjs}
The flexible partitioning mechanism described in Section~\ref{sec:partitioning} introduces a significant challenge: workload heterogeneity. Since owners select different cut points $l_n$ based on their distinct privacy-cost trade-offs, the resulting tasks exhibit high variance in their resource consumption profiles (i.e., varying ratios of communication to computation). This heterogeneity, compounded by the sequential dependency of collaborative tasks, often causes critical resources to remain idle while others are blocked (e.g., the server sits idle waiting for large uploads). To address this inefficiency and maximize global resource utilization as well as system throughput, we propose the PDJS algorithm.


\subsubsection{Discrete Event System Modeling}
We model the NWaaS system as a three-stage tandem queue system (formally known as a flow shop \cite{garey1976complexity}) consisting of three shared, constrained resources: uplink bandwidth ($R^{\text{up}}$), server computation capacity ($\omega_s$), and downlink bandwidth ($R^{\text{down}}$). However, unlike traditional tandem queues where tasks are often independent and single-pass, the collaborative nature of NWaaS introduces a complex, iterative interaction pattern. Each model owner $o_n$ requires multiple rounds of back-and-forth communication (epochs and batches) with the server. Consequently, the watermarking tasks exhibit strong sequential interactive dependencies, distinguishing our problem from standard independent job scheduling \cite{garey1976complexity}. Note that the client-side local computation is owner-exclusive and does not contend for system resources.

For owner $o_n$, the watermarking process involves $E_n$ epochs with $B_n$ batches per epoch. Let $K_n = E_n \times B_n$ denote the total number of task units (batches) for owner $o_n$. The total system workload is a collection of tasks $\mathcal{T} = \{ \tau_{n,k} \mid 1 \le n \le N, 1 \le k \le K_n \}$.
Based on the cost profile $(\eta_n, A_n, G_n)$ derived in Section~\ref{sec:partitioning}, the processing time for a single task unit $\tau_{n,k}$ on the three resources is defined as:
\begin{equation}
    t_n^{\text{up}} = \frac{A_n}{R_n^{\text{up}}}, \quad t_n^{\text{srv}} = \frac{\eta_n C_n}{\omega_s}, \quad t_n^{\text{down}} = \frac{G_n}{R_n^{\text{down}}},
\end{equation}
where $R_n^{\text{up}}$ and $R_n^{\text{down}}$ are the effective data transmission rates for owner $o_n$. Note that unlike the coarse estimation $B_{\text{est}}$ used for local partitioning in (\ref{eq:local_opt}), the scheduler operates on these actual, often asymmetric rates. The local computation time $t_n^{\text{loc}} = \Phi_n(l_n)/\omega_n$ acts as a prerequisite delay before the task enters the uplink queue.

\begin{algorithm}[t]
    \caption{Proportion Disparity Joint Scheduling (PDJS)}
    \label{alg:pdjs}
    \begin{algorithmic}[1]
        \STATE \textbf{Input:} Owner profiles $\{(\eta_n, A_n, G_n)\}_{n=1}^N$, Bandwidth estimates.
        \STATE \textbf{Initialize:} System Load Vector $\boldsymbol{\lambda}_{\text{sys}} \leftarrow [0, 0, 0]$.
        \STATE \textbf{While} there are unfinished tasks \textbf{do}
        \STATE \hspace{0.5cm} Update queues $\mathcal{Q}_U, \mathcal{Q}_S, \mathcal{Q}_D$ based on task completions.
        \STATE \hspace{0.5cm} Update $\boldsymbol{\lambda}_{\text{sys}}$ according to Eq.~\eqref{eq:buf_prop}.
        \STATE \hspace{0.5cm} \textbf{If} Uplink Resource is idle \textbf{then}
        \STATE \hspace{1.0cm} Identify candidate tasks $\mathcal{T}_{\text{ready}}$ (local compute finished).
        \STATE \hspace{1.0cm} \textbf{For} each task $\tau_z \in \mathcal{T}_{\text{ready}}$ \textbf{do}
        \STATE \hspace{1.5cm} Calculate task profile $\boldsymbol{\mu}_z$.
        \STATE \hspace{1.5cm} Compute disparity $d_z$ via Eq.~\eqref{eq:prop_dist}.
        \STATE \hspace{1.0cm} \textbf{End For}
        \STATE \hspace{1.0cm} Schedule task $\tau_{z^*} \leftarrow \arg \max d_z$ to Uplink via Eq.~\eqref{eq:pdjs}.
        \STATE \hspace{0.5cm} \textbf{End If}
        \STATE \hspace{0.5cm} (Similar scheduling logic for Server and Downlink threads)
        \STATE \textbf{End While}
    \end{algorithmic}
\end{algorithm}

Let $\Pi$ denote the global scheduling strategy. Accordingly, let $C_i^{\text{up}}, C_i^{\text{srv}}, C_i^{\text{down}}$ denote the completion timestamps of the $i$-th task in the respective pipelines determined by $\Pi$. Due to the serial dependency (Uplink $\to$ Server $\to$ Downlink), the completion time is recursive. For instance, the server computation for the $i$-th task (belonging to owner $o_{n(i)}$) cannot start until both the previous server task is finished and the current task's data has been uploaded:
\begin{equation}
\label{eq:pipeline_dependency}
    C_i^{\text{srv}} = \max\left(C_{i-1}^{\text{srv}}, \ C_{j(i)}^{\text{up}}\right) + t_{n(i)}^{\text{srv}},
\end{equation}
where $C_{j(i)}^{\text{up}}$ denotes the uplink completion time of the specific task unit that is currently scheduled as the $i$-th task on the server, and $j(i)$ represents its corresponding index in the uplink sequence. Similar dependencies apply to the server and downlink phase.

The objective of the scheduler is to determine the optimal ordering to maximize the system throughput, which is equivalent to minimize the average service completion time (makespan) across all $N$ owners:
\begin{equation}
\label{eq:scheduling_obj}
    \min_{\Pi} \quad \frac{1}{N} \sum_{n=1}^N \mathcal{D}_{e(n)}^{\text{down}},
\end{equation}
where $\mathcal{D}_n^{\text{down}}$ is the timestamp when owner $o_n$ receives the final gradient of the last batch and $e(n)$ is the corresponding task index.

\subsubsection{Adaptive Resource Allocation Strategy}
Problem~\eqref{eq:scheduling_obj} corresponds to the classic three-machine flow shop scheduling, and finding its global optimum is known to be NP-hard \cite{garey1979computers}. Although generic heuristics exist for such problems, they fail to exploit the unique workload heterogeneity in NWaaS, where task profiles vary drastically due to different owner-defined privacy settings. Moreover, simple strategies like First-In-First-Out (FIFO) are inefficient here because of the head-of-line blocking phenomenon. For example, in an uplink-constrained scenario, an owner with a shallow-offload strategy (massive $A_n$, small $\eta_n$) might monopolize the uplink channel. During this time, the server GPU and downlink channel remain starved, resulting in low resource utilization and extended processing time.

To address this, PDJS employs a greedy strategy based on payload balancing. The core intuition is to prioritize tasks whose resource demands inversely match the current system congestion profile. If the uplink is congested, we should schedule a task that requires minimal uplink but heavy server computation, effectively filling the gaps in the resource timeline.

\begin{table*}[!t]
  \centering
  \renewcommand{\arraystretch}{1.3}
  \caption{Summary of implemented tasks and models.}
\label{tab:summary_implementation}
  \resizebox{1\textwidth}{!}{
  \begin{tabular}{c|c|c|c|c|c|c|c|c|c}
  \hline
  \hline
  \textbf{Type}$^\dag$ & \textbf{Task} & \textbf{Dataset} & $\mathbb{M}_\theta$ & \textbf{Backbone} & $\mathbb{G}_\gamma$ & $\mathbb{D}_\delta$ & $\mathtt{dim}(\mathsf{k})$ & $\mathtt{dim}(x)$ &  $\mathtt{dim}(\mathbb{M}_\theta(x))^\ddag$\\
  \hline

  \multirow{5}{*}{I2I} & Dehazing & NYU Depth V2 \cite{Silberman:ECCV12_Dataset_NYU} & AODnet \cite{li2017aod_Victim_AOD} & \multirow{3}{*}{CNN} & \multirow{5}{*}{\begin{tabular}[c]{@{}c@{}}CGAN\\ Generator\end{tabular}} & \multirow{8}{*}{{\begin{tabular}[c]{@{}c@{}}CEILNet\\ \cite{fan2017generic}\end{tabular}}} & \multirow{7}{*}{$(1,256)$} & $(3, 480, 640)$& $(3, 480, 640)$\\
    \cline{2-4} \cline{9-10}
    
  & Style Trans. &  \begin{tabular}[c]{@{}c@{}}PASCAL VOC \cite{everingham2010pascal_Dataset_Pascal}\\WikiArt \cite{artgan2018_Dataset_Wikiart}\end{tabular}  & LinearTransfer \cite{li2018learning_Victim_StyleTransfer} & & & & & \multirow{2}{*}{$(3, 256, 256)$}  & \multirow{2}{*}{$(3, 256, 256)$}\\ \cline{2-5}

  & Inpainting & CelebA \cite{liu2015faceattributes_Dataset_CelebA} & MAT \cite{li2022mat_Victim_MAT} & \multirow{3}{*}{Transformer} & & & & & \\
  \cline{2-4} \cline{9-10}

  & Denoising & CDD-11 \cite{guo2024onerestore_Victim} & OneRestore \cite{guo2024onerestore_Victim} & & & & & $(3, 224, 224)$ & $(3, 224, 224)$\\ \cline{2-4} \cline{9-10}

  & Super Res. & PASCAL VOC \cite{everingham2010pascal_Dataset_Pascal} & SwinIR \cite{liang2021swinir_Victim} & & & & & $(3, 256, 256)$& $(3, 512, 512)$\\\cline{1-6} \cline{9-10}

  N2I & \multirow{2}{*}{Digit Gen.} & \multirow{3}{*}{Data-Free} & GAN \cite{goodfellow2014generativeadversarialnetworks_VictimGAN} & \multirow{3}{*}{CNN}& \multirow{2}{*}{MLP}& & & {$(1, 100)$}&\multirow{2}{*}{$(1, 256, 256)$} \\
  \cline{1-1}  \cline{4-4} \cline{9-9}

  NT2I & & & CGAN \cite{mirza2014conditional_Victim_ConditionalGAN} & & & & & $(1, 110)$& \\
  \cline{1-2} \cline{4-4} \cline{6-6} \cline{8-10} 

  T2I & Image Gen. & & Stable Diffusion \cite{rombach2022high_Victim_StableDiffusion}& & Upsampler & & $(1,77)$ & $(77, 768)$ & $(3, 512, 512)$\\ 
  \hline
  \hline
  \multicolumn{10}{l}{$\dag$ I2I: Image-to-Image. N2I: Noise-to-Image. NT2I: Noise-and-Text-to-Image. T2I: Text-to-Image. }\\
  \multicolumn{10}{l}{$\ddag$ Size of watermark $m$ is by default the same as the model output, i.e., $\mathtt{dim}(m) = \mathtt{dim}(\mathbb{M}_\theta(x))$.}
  \end{tabular}}
\end{table*}

\textbf{State Representation:}
At any scheduling instant $t$, let $\mathcal{Q}_U, \mathcal{Q}_S, \mathcal{Q}_D$ be the set of tasks currently queued in the buffers of the Uplink, Server, and Downlink nodes, respectively. We define the \textit{system load vector} $\boldsymbol{\lambda}_{\text{sys}}$ as the aggregate pending workload in the buffers:
\begin{equation}
\label{eq:buf_prop}
    \boldsymbol{\lambda}_{\text{sys}} = \left[ \sum_{k \in \mathcal{Q}_U} t_{n(k)}^{\text{up}}, \quad \sum_{k \in \mathcal{Q}_S} t_{n(k)}^{\text{srv}}, \quad \sum_{k \in \mathcal{Q}_D} t_{n(k)}^{\text{down}} \right].
\end{equation}
This vector reflects the current bottlenecks. For a candidate task $\tau_z$ from owner $o_z$, its task profile vector is $\boldsymbol{\mu}_z = [t_z^{\text{up}}, t_z^{\text{srv}}, t_z^{\text{down}}]$.

\textbf{Proportion Disparity Metric:}
We normalize both vectors to represent the \textit{proportion} of resource occupancy. The suitability of scheduling task $\tau_z$ is measured by how much its profile differs from the current congestion profile. We define the proportion disparity distance $d_z$ using the $L_1$-norm:
\begin{equation}
\label{eq:prop_dist}
    d_z = \left\| \frac{\boldsymbol{\lambda}_{\text{sys}}}{\|\boldsymbol{\lambda}_{\text{sys}}\|_1} - \frac{\boldsymbol{\mu}_z}{\|\boldsymbol{\mu}_z\|_1} \right\|_1.
\end{equation}
A larger $d_z$ implies that the task $\tau_z$ demands resources that are currently strictly underutilized by the system (i.e., distinct from the congested resource).

\textbf{Algorithm Execution:}
The PDJS algorithm operates dynamically. Whenever a resource becomes free, the scheduler evaluates all candidate tasks in the waiting pool. It selects the task $\tau_{z^*}$ that maximizes the disparity $d_z$:
\begin{equation}
\label{eq:pdjs}
    \tau_{z^*} = \arg \max_{\tau_z \in \mathcal{T}_{\text{ready}}} d_z.
\end{equation}
By consistently selecting tasks that balance out the load vector, PDJS ensures that all three resources (uplink, server, downlink) are kept active simultaneously, thereby maintaining near-optimal resource utilization. The detailed procedure is summarized in Algorithm~\ref{alg:pdjs}.

\section{Experimental Evaluation}
In this section, we evaluate the proposed NWaaS framework to validate its robustness, security, efficiency, and system reliability. Since our non-intrusive paradigm targets zero performance degradation and privacy-preserving partitioning, which are properties absent in existing intrusive methods, direct comparisons with traditional baselines are inapplicable. Therefore, the experiments comprehensively cover the effectiveness of the $\mathtt{ShadowMark}$ algorithm across diverse settings and the performance of the PDJS scheduling in collaborative environments.

\subsection{Experiment Settings}
\subsubsection{Implementation Details}
For $\mathtt{ShadowMark}$, as summarized in Table~\ref{tab:summary_implementation}, we implement $\mathtt{ShadowMark}$ across four diverse X-to-Image tasks: Image-to-Image (I2I), Noise-to-Image (N2I), Noise-and-Text-to-Image (NT2I), and Text-to-Image (T2I). To verify broad applicability, our evaluation encompasses a wide range of datasets (NYU Depth V2, WikiArt, CelebA, etc.) and target model architectures ($\mathbb{M}_\theta$), ranging from standard CNNs to Transformers and Foundation Models (e.g., Stable Diffusion \cite{rombach2022high_Victim_StableDiffusion}). Furthermore, we select different watermarking modules ($\mathbb{G}_\gamma$ and $\mathbb{D}_\delta$) with varying parameter sizes and architectures to demonstrate the adaptability and parameter-efficiency for training. These modules are optimized using the Adam optimizer with an initial learning rate of $2 \times 10^{-4}$ under a global random seed of $44$. To evaluate key security, the number of random guesses is set to $1$ million times in each experiment.

\begin{figure*}[!t]
    \centering
    \includegraphics[width=1.743\columnwidth]{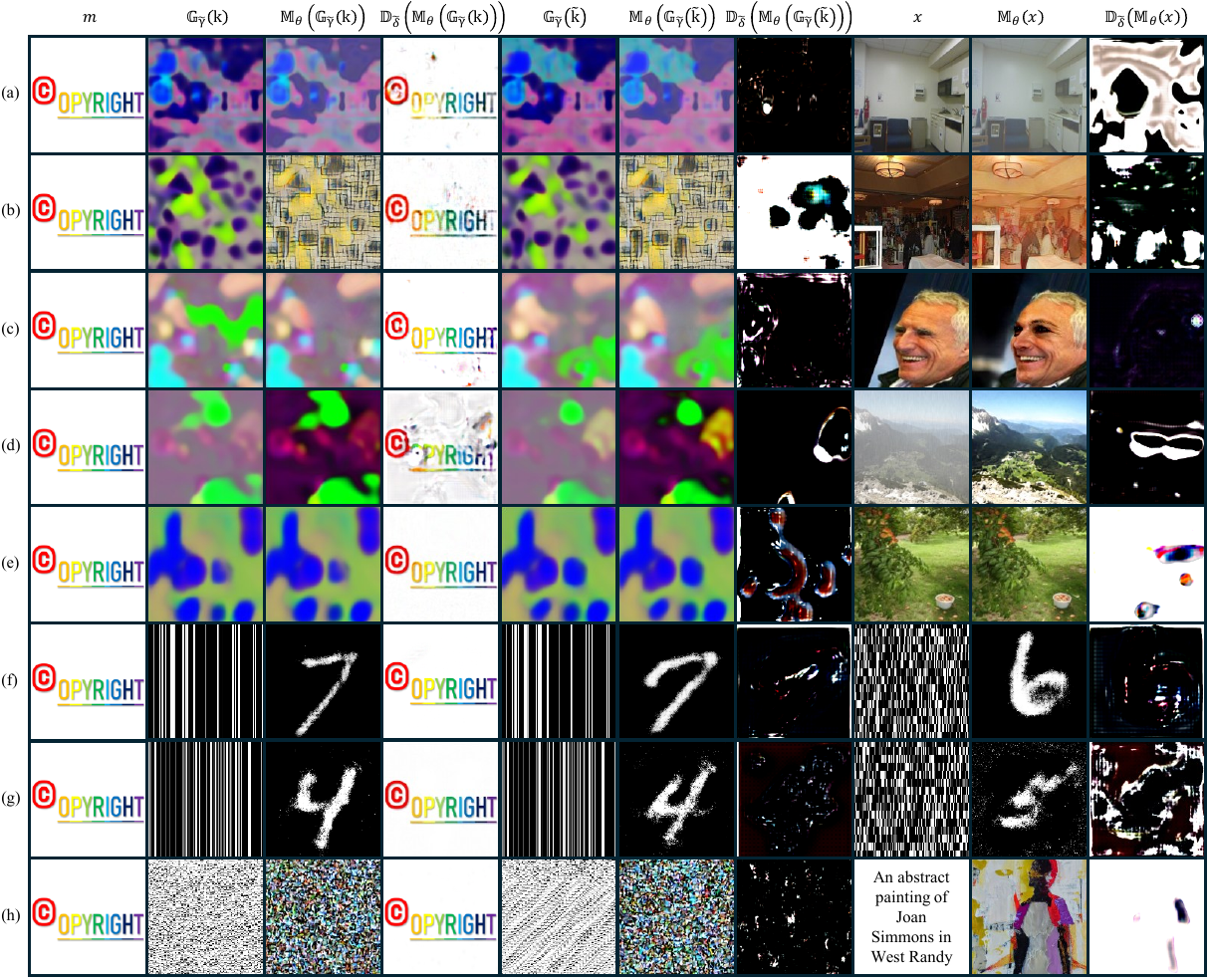}
    \caption{Qualitative results of $\mathtt{ShadowMark}$ for (a)-(e) I2I (AODnet \cite{li2017aod_Victim_AOD}, LinearTransfer \cite{li2018learning_Victim_StyleTransfer}, MAT \cite{li2022mat_Victim_MAT}, OneRestore \cite{guo2024onerestore_Victim}), and SwinIR \cite{liang2021swinir_Victim}), (f) N2I (GAN \cite{goodfellow2014generativeadversarialnetworks_VictimGAN}), (g) NT2I (CGAN \cite{mirza2014conditional_Victim_ConditionalGAN}), and (h) T2I (Stable Diffusion \cite{rombach2022high_Victim_StableDiffusion}).}
    \label{fig:qualitative_shadowmark}
\end{figure*}

\begin{figure}[!t]
    \centering
    \includegraphics[width=\columnwidth]{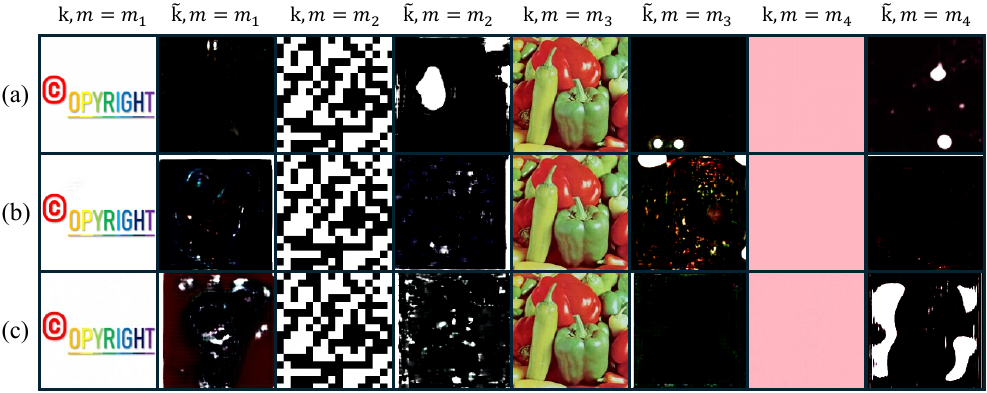}
    \caption{Sensitivity of $\mathtt{ShadowMark}$ to $4$ choices of $m$ for (a) I2I (AODnet \cite{li2017aod_Victim_AOD}), (b) N2I (GAN \cite{goodfellow2014generativeadversarialnetworks_VictimGAN}), and (c) NT2I (CGAN \cite{mirza2014conditional_Victim_ConditionalGAN}), which shows the extracted $\hat{m}$ with the correct key $\mathsf{k}$ and wrong key $\tilde{\mathsf{k}}$ for each mark $m_i$.}
    \label{fig:watermark_sensitivity}
\end{figure}

To evaluate the performance of PDJS in balancing resource allocation and its reliability, we compare the training duration and resource efficiency under two distinct categories of scenarios: resource-limited and resource-balanced. 
The resource-limited scenarios represent conditions where the system bottleneck shifts among three critical resources: uplink communication resource (UCR), server computation resource (SCR), and downlink communication resource (DCR). Specifically, for each bottleneck scenario, we designate $80\%$ of the clients as heavy consumers of the respective target resource (e.g., having large upload payloads in the UCR-limited case), thereby creating significant congestion at the bottleneck. In contrast, the resource-balanced scenario maintains an equivalent proportion of clients across different resource occupation profiles. We employ an FIFO scheduling strategy as the baseline for comparison. The system is deployed on a server cluster equipped with 4 NVIDIA RTX 5000 Ada Generation GPUs (as shown in Fig.~\ref{fig:PDJS_performance}(a)). In our experiments, the proposed collaborative partitioning mechanism is operationalized by randomly assigning heterogeneous resource demand profiles to clients (subject to the specific resource scenarios), effectively simulating the varying communication loads ($A_n, G_n$) and computation loads ($\eta_n$) resulting from different owner-selected cut points $l_n$.

\subsubsection{Metrics}
We employ three key metrics to quantify verifiability and security:
\begin{itemize}
    \item \textbf{Ownership Verification:} For the authorized model $\mathbb{M}_\theta$, we use NCC with the criterion defined in (\ref{eq:verify_m}).
    \item \textbf{Model Stealing Tracing:} For verifying suspected surrogate models $\mathbb{S}_{\tilde{\vartheta}}$, which may suffer from domain shifts, we use NCCD as defined in (\ref{eq:verify_s}).
    \item \textbf{Security:} To quantify watermark security against brute-force attacks, we define the success rate of ambiguity ($\mathtt{SR}_\text{A}$) as the number of discovered valid keys divided by the total number of random trials.
\end{itemize}

\subsection{Qualitative Evaluation}
\label{sec:qualitative_val}

\subsubsection{$\mathtt{ShadowMark}$ Procedure} \label{sec:shadowmark_visual}

We first verify the feasibility of $\mathtt{ShadowMark}$ by demonstrating that the owner-defined watermark image $m$ can be extracted from the protected model with the model strictly untouched. Fig.~\ref{fig:qualitative_shadowmark} presents the qualitative results of $\mathtt{ShadowMark}$ across $4$ tasks using models with $x$ from $6$ datasets. The columns in the figure, from left to right, are watermark $m$, encoded correct key $\mathbb{G}_{\tilde{\gamma}}(\mathsf{k})$, processed encoded correct key by $\mathbb{M}_\theta$, extracted mark with correct key $\mathbb{D}_{\tilde{\delta}}(\mathbb{M}_{{\theta}}(\mathbb{G}_{\tilde{\gamma}}(\mathsf{k})))$, encoded wrong key $\mathbb{G}_{\tilde{\gamma}}(\mathsf{\tilde{k}})$, processed encoded wrong key by $\mathbb{M}_\theta$, extracted mark with wrong key $\mathbb{D}_{\tilde{\delta}}(\mathbb{M}_{{\theta}}(\mathbb{G}_{\tilde{\gamma}}(\mathsf{\tilde{k}})))$, normal query $x$, normal query output $\mathbb{M}_\theta$, and decoded mark from $x$, i.e., $\mathbb{D}_{\tilde{\delta}}\left(\mathbb{M}_{{\theta}}\left(x\right)\right)$. It can be seen that for all tested models, $\mathtt{ShadowMark}$ successfully extracts the watermark when provided with the correct key, while producing meaningless patterns when the key is incorrect. We notice that the encoded keys generated from an incorrect key, i.e., $\mathbb{G}_{\tilde{\gamma}}(\mathsf{\tilde{k}})$ can be visually similar to that generated from the correct key, i.e., $\mathbb{G}_{\tilde{\gamma}}(\mathsf{k})$, but the extracted outputs are significantly different. Note that these intermediate results are private and secured in the black-box API of $\mathtt{ShadowMark}$.

\subsubsection{Sensitivity to Choice of Watermark $m$}
\label{sec:different_watermarks}
We now illustrate the sensitivity of $\mathtt{ShadowMark}$ to the choices of $m$, and the results are presented in Fig.~\ref{fig:watermark_sensitivity}. Specifically, we test $4$ types of $m$ (denoted by $m_i$) including ``COPYRIGHT'' text image $m_1$, binary pattern $m_2$, color image $m_3$, and pure color $m_4$. The subfigures present the results for $3$ different tasks, when the correct key $\mathsf{k}$ and an incorrect key $\tilde{\mathsf{k}}$ is used, respectively. It can be seen that for all cases, 
$\mathtt{ShadowMark}$ can successfully extract the watermark with the correct key while generating meaningless patterns for incorrect keys, which verifies that $\mathtt{ShadowMark}$ can extract arbitrary owner-defined watermark $m$ without modifying the protected model.

\begin{table*}[!t]
\centering
\caption{Quantitative results of $\mathtt{ShadowMark}$, where $\mathtt{NCC}$ and $\mathtt{NCCD}$ are evaluated using (\ref{eq:verify_m}) and (\ref{eq:verify_s}), and $0<\mathtt{SR}_{\text{A}}<1$.}
\resizebox{1.0\textwidth}{!}{
\begin{tabular}{c|c|c|c|c|c|c|c|c|c}
\hline
\hline
\textbf{Index} & \textbf{Type} & $\mathbb{M}_\theta$ & $\mathbb{G}_\gamma$ & $\mathbb{D}_\delta$  & $\mathbb{S}_{{\vartheta}}$ & $m$ & $\mathtt{NCC}$ & $\mathtt{NCCD}$ & $\mathtt{SR}_{\text{A}}$ \\ \hline
1 & \multirow{10}{*}{I2I} & \multirow{6}{*}{AODnet (0.002M) \cite{li2017aod_Victim_AOD}} & \multirow{2}{*}{UNet (41.83M) \cite{ronneberger2015u_UNet}} & CEILNet (3.37M) \cite{fan2017generic} & \multirow{6}{*}{UNet} & \multirow{11}{*}{``COPYRIGHT''} & $0.99$ & $0.61$  & $0.0$ \\  \cline{5-5}
2 &  &  &  & EEENet (0.12M) \cite{Wu2021Watermarking_Box_Free} &  &  & $0.99$ & $0.60$ & $0.0$ \\  \cline{4-5}
3 &  &  & \multirow{2}{*}{\begin{tabular}[c]{@{}c@{}}ResNet Generator \\ (12.71M) \end{tabular}} & CEILNet (3.37M) \cite{fan2017generic} &  &  & $0.99$ & $0.67$ & $0.0$ \\ \cline{5-5}
4 &  &  &  & EEENet (0.12M) \cite{Wu2021Watermarking_Box_Free}&  &  & $0.99$ & $0.70$ & $0.0$ \\  \cline{4-5}
5 &  &  & \multirow{6}{*}{\begin{tabular}[c]{@{}c@{}}CGAN Generator \\ (17.36M) \end{tabular}} & CEILNet (3.37M) \cite{fan2017generic} &  &  & $0.99$ & $0.71$  & $0.0$ \\  \cline{5-5}
6 &  &  &  & EEENet (0.12M) \cite{Wu2021Watermarking_Box_Free} &  &  & $0.99$ & $0.65$ & $0.0$  \\  \cline{3-3} \cline{5-6} \cline{8-10}
7 &  & LinearTransfer (12.156M) \cite{li2018learning_Victim_StyleTransfer} &  & \multirow{10}{*}{CEILNet (3.37M) \cite{fan2017generic}} & -- &  & $0.99$ & -- & $0.0$ \\  \cline{3-3} \cline{6-6}
8 &  & MAT (59.78M) \cite{li2022mat_Victim_MAT} &  &  & -- &  & $0.99$ & -- & $0.0$  \\  \cline{3-3} \cline{6-6}
9 &  & OneRestore (18.12M) \cite{guo2024onerestore_Victim} &  &  & UNet &  & $0.86$ & $0.84$ & $0.0$  \\\cline{3-3} \cline{6-6}
10 &  & SwinIR (11.75M) \cite{liang2021swinir_Victim}&  &  & -- &  & $0.99$ & -- & $0.0$ \\ \cline{1-4} \cline{6-6} \cline{8-10}
11 & \multirow{4}{*}{N2I} & \multirow{4}{*}{GAN (67.88M) \cite{goodfellow2014generativeadversarialnetworks_VictimGAN}} & \multirow{5}{*}{FCN (0.05M)}  &  & \multirow{5}{*}{CNN} &  & $0.99$ & $0.60$ & $0.0$ \\  \cline{7-7}
12 &  &  &  &  &  & Binary & $0.99$ & $0.10$ & $0.0$ \\  \cline{7-7}
13 &  &  &  &  &  & Pepper & $0.99$ & $0.04$ & $0.0$ \\  \cline{7-7}
14 &  &  &  &  &  & Pink & $0.99$ & $0.88$ & $0.0$ \\ \cline{1-3} \cline{7-10}
15 & NT2I & CGAN (67.88M) \cite{mirza2014conditional_Victim_ConditionalGAN} & &  &  & \multirow{2}{*}{``COPYRIGHT''} & $0.99$ & $0.70$ & $0.0$ \\ \cline{1-4} \cline{6-6} \cline{8-10}
16 & T2I & Stable Diffusion (859.52M) \cite{rombach2022high_Victim_StableDiffusion}& Upsampler (2.36M) &  & -- &  & $0.99$ & -- & $0.0$ \\ \hline\hline
\end{tabular}}
\label{tab:quantitative_table}
\end{table*}

\begin{figure*}[!t]
    \centering
    \includegraphics[width=2.0\columnwidth]{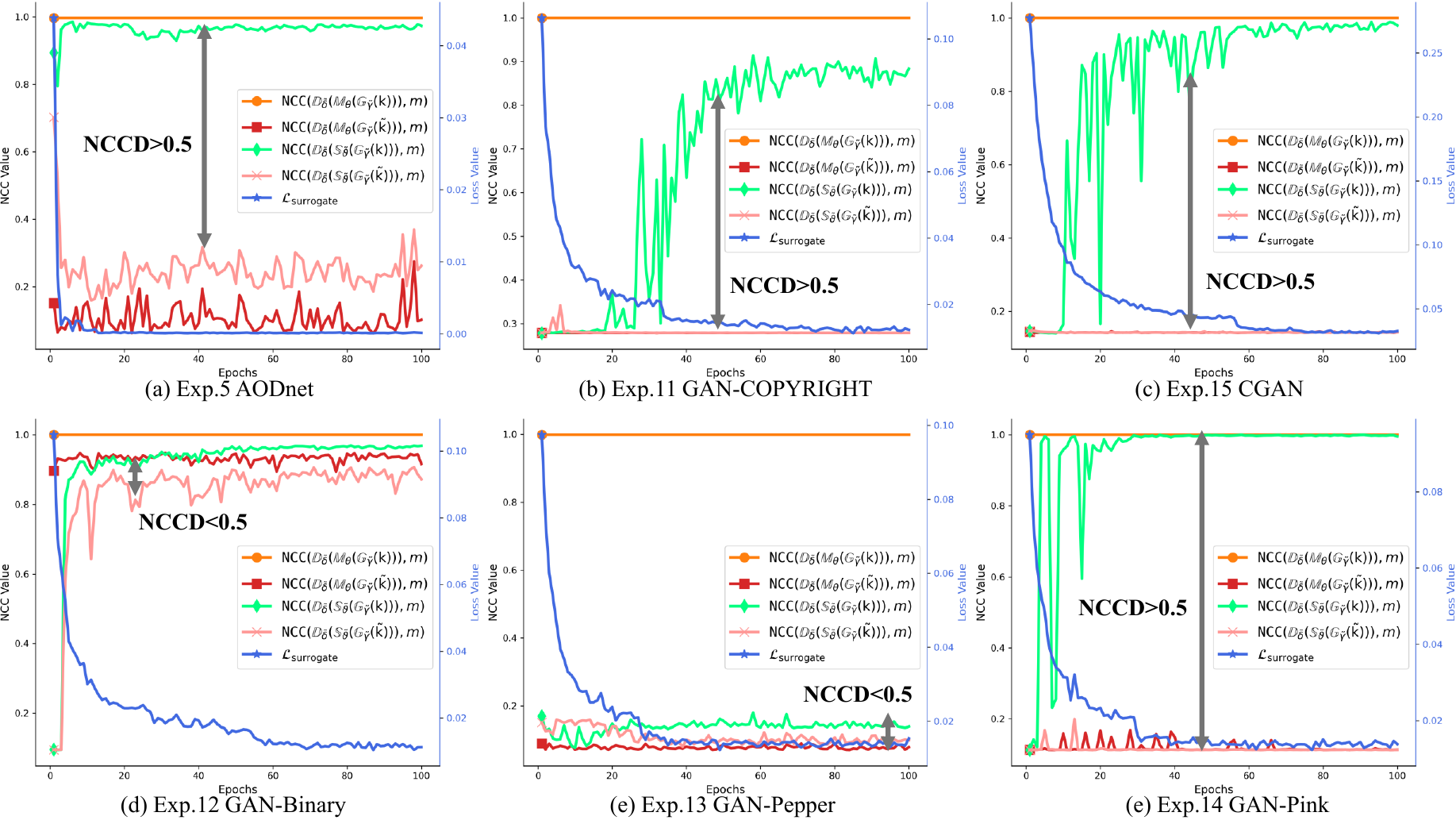}
        \caption{Loss and $\mathtt{NCC}$ values versus epoch during surrogate training in surrogate training for $6$ experiments in Table \ref{tab:quantitative_table}.}
    \label{fig:line_chart_surrogate}
\end{figure*}

\begin{figure}[!t]
    \centering
    \includegraphics[width=1.0\columnwidth]{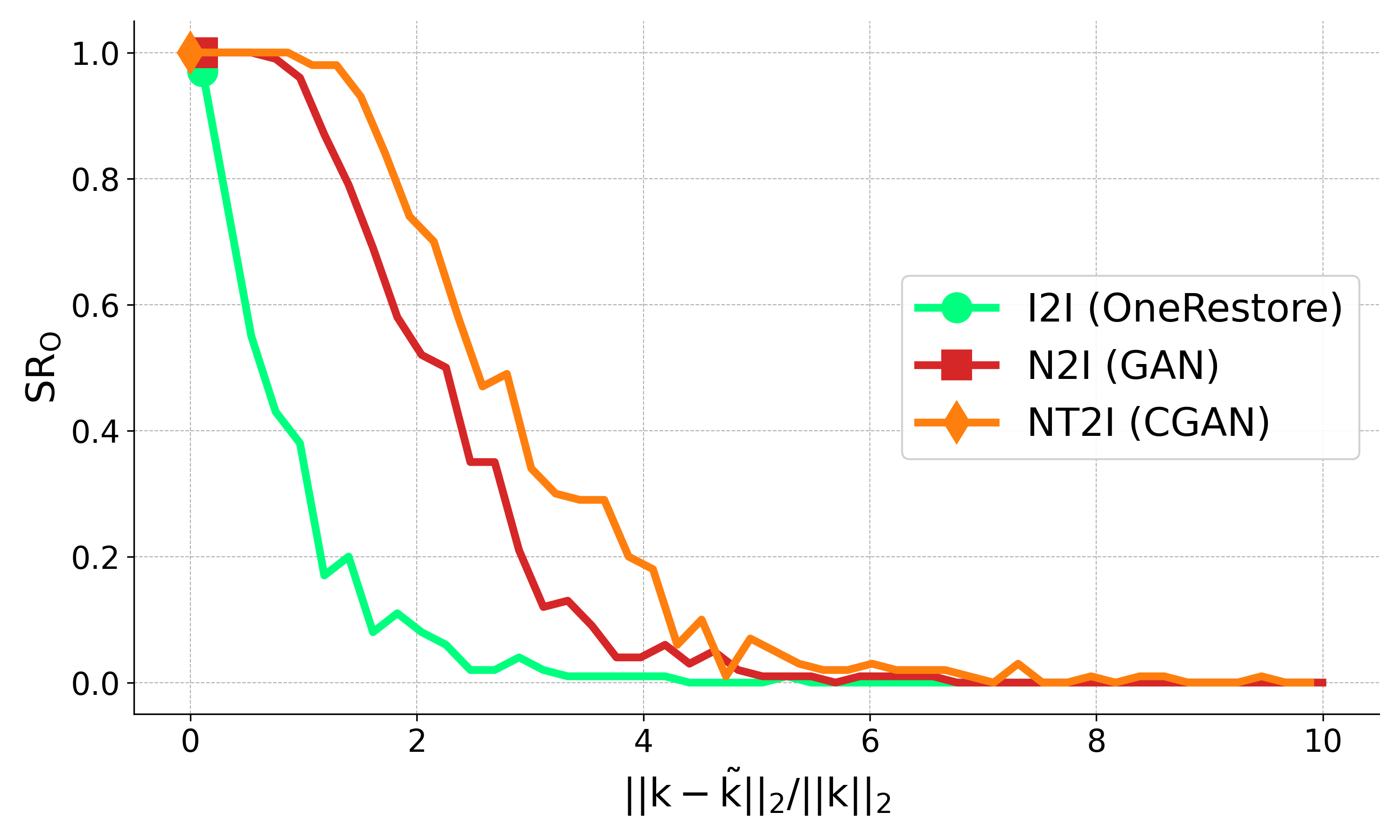}
    \caption{Key security test results for $3$ experiments, where $\mathsf{dim}(\mathtt{k}) = (1,256)$.}
    \label{fig:radius_security}
\end{figure}

\begin{figure*}[!t]
    \centering
    \includegraphics[width=2.0\columnwidth]{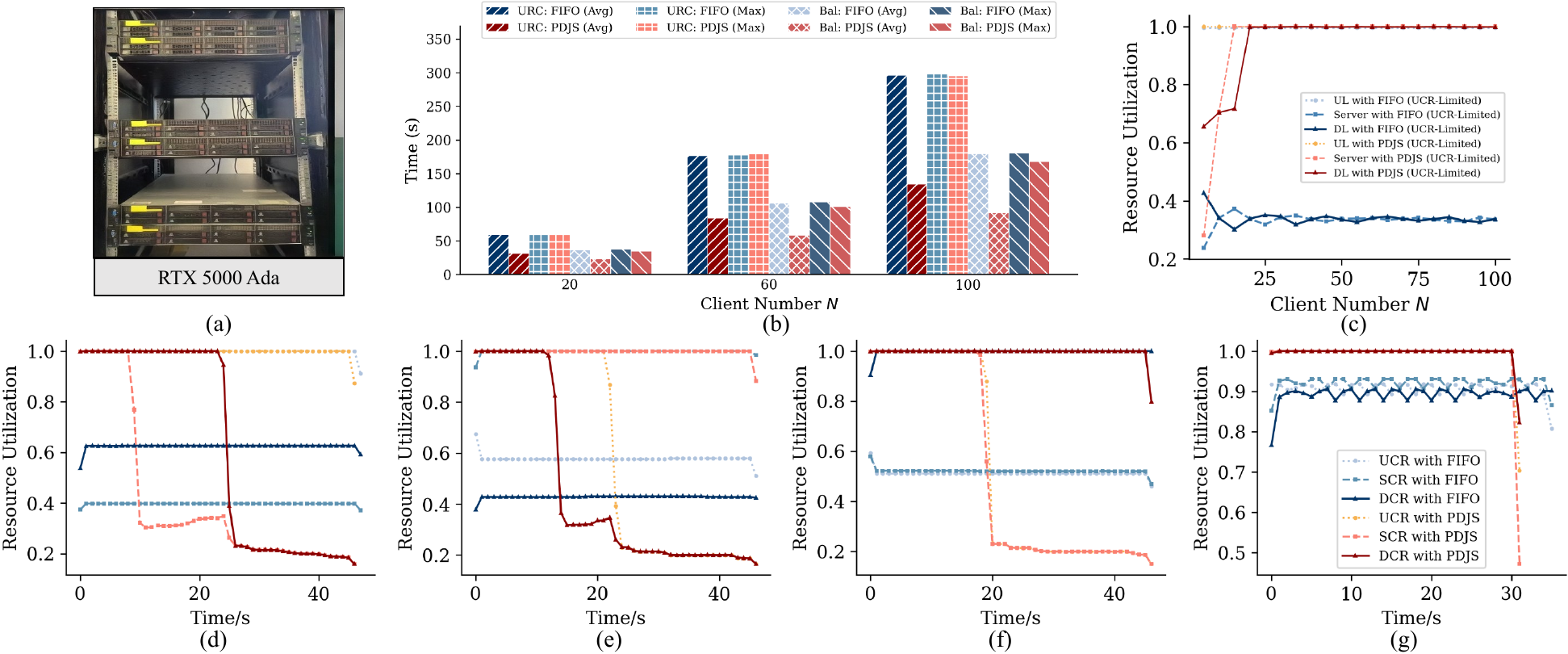}
        \caption{Evaluation of the proposed PDJS algorithm. (a) Deployed server cluster equipped with four NVIDIA RTX 5000 Ada GPUs. (b) Comparison of average and maximum service completion times between FIFO and PDJS strategies across the varying number of concurrent clients under URC-limited and resource-balanced scenarios. (c) Resource utilization rates versus client volume under the URC-limited scenario. (d)–(g) Real-time resource utilization under specific bottleneck scenarios: (d) URC-limited, (e) SCR-limited, (f) DCR-limited, and (g) resource-balanced. The legend in (g) applies to sub-figures (d)–(g).}
    \label{fig:PDJS_performance}
\end{figure*}

\subsection{Quantitative Evaluation}
We now present the qualitative experimental results to further verify the performance of $\mathtt{ShadowMark}$. Based on the task and model summary in Table \ref{tab:summary_implementation}, we have carried out $16$ different experiments spanning I2I, N2I, NT2I, and T2I, for watermark verification in both the original and surrogate models, as well as the brute force key ambiguity attack. The results are presented in Table \ref{tab:quantitative_table}, where the key length and dataset information are the same as those in Table \ref{tab:summary_implementation} and omitted.

\begin{table}[!t]
\centering
\caption{$\mathtt{NCC}$ for Different Epochs of OneRestore Model under Valid Key}
\resizebox{0.95\columnwidth}{!}{
\begin{tabular}{c|c|c}
\hline\hline
\textbf{Checkpoint} & \textbf{Dataset} & \textbf{$\mathtt{NCC}$ (Valid Key)} \\ \hline
Epoch 0 & CDD-11 & $0.231$ \\ \hline
Epoch 10 & CDD-11 & $0.134$ \\ \hline
Epoch 30 & CDD-11 & $0.184$ \\ \hline
Epoch 40 & CDD-11 & $0.559$ \\ \hline
Epoch 50 & CDD-11 & $0.678$ \\ \hline
Epoch 60 & CDD-11 & $0.558$ \\ \hline
Epoch 70 & CDD-11 & $0.629$ \\ \hline
\textbf{Epoch 80} & \textbf{CDD-11} & $\mathbf{0.866}$ \\ \hline
Epoch 90 & CDD-11 & $0.559$ \\ \hline
\textbf{Epoch 80} & \textbf{RESIDE \cite{li2018benchmarking}} & $\mathbf{0.458}$ \\ \hline\hline
\end{tabular}}
\label{tab:rebuttal_near_neighbor}
\end{table}

\subsubsection{Watermark Verification in the Original Model}
 For the original model verification, where the threshold is $0.95$ according to (\ref{eq:verify_m}), we observe that $\mathtt{NCC}=0.99 > 0.95$ for $15$ out of $16$ experiments, while only in Task $9$ when protecting OneRestore~\cite{guo2024onerestore_Victim}, $\mathtt{NCC}=0.86 < 0.95$. It can be seen from Fig.~\ref{fig:qualitative_shadowmark} (d) that the extracted mark from OneRestore, i.e., $\mathbb{D}_{\tilde{\delta}}(\mathbb{M}_{{\theta}}(\mathbb{G}_{\tilde{\gamma}}(\mathsf{k})))$, has relatively the lowest visual quality, but it still resembles the ``COPYRIGHT'' information. We note that OneRestore is based on Transformer, but this problem does not exist for the other $2$ Transformer-based models, i.e., MAT \cite{li2022mat_Victim_MAT} and SwinIR \cite{liang2021swinir_Victim}. This indicates that although $\mathtt{ShadowMark}$ is applicable to diverse types of DNNs, watermarking performance can be affected by the specific architecture of the protected model.

 Additionally, we evaluate the specificity of $\mathtt{ShadowMark}$ using near-neighbor checkpoints of OneRestore and an independently trained model on RESIDE \cite{li2018benchmarking}, with the quantitative results reported in Table~\ref{tab:rebuttal_near_neighbor}. The independently trained model yields a low $\mathtt{NCC}$ of $0.458$, firmly ruling out false positives. For sibling checkpoints from the identical training pipeline, early epochs yield minimal responses ($\mathtt{NCC} < 0.24$), while late-stage sibling checkpoints (Epochs $40$--$70$, $90$) exhibit moderately elevated responses. This suggests that while strict authorization is exclusively granted to the exact target checkpoint (Epoch 80, $\mathtt{NCC}=0.866$), the partial manifold overlap enables the tracing of unauthorized leakages of intermediate sibling models.

\subsubsection{Robustness Against Model Extraction}
We note that since the protected model is deployed via a black-box API, traditional parameter-level attacks (e.g., fine-tuning or pruning) are infeasible. Furthermore, because $\mathtt{ShadowMark}$ is strictly non-intrusive, normal API outputs do not carry the watermark, rendering conventional post-processing attacks (e.g., JPEG compression, noise addition, or regeneration attack on outputs) completely ineffective. Therefore, the primary threat is the model extraction attack, which has the following properties: (1) surrogate training can approximate the functionality of $\mathbb{M}_\theta$ with controlled degradations and (2) the cost of surrogate training should be smaller than training the model from the scratch without surrogate data. Otherwise, model extraction attack is considered infeasible. We found that LinearTransfer (style transfer) \cite{li2018learning_Victim_StyleTransfer}, MAT (super-resolution) \cite{li2022mat_Victim_MAT}, SwinIR (inpainting) \cite{liang2021swinir_Victim}, and Stable Diffusion (image generation) \cite{rombach2022high_Victim_StableDiffusion} did not meet the two conditions and their corresponding $\mathtt{NCCD}$ values are invalid in Table \ref{tab:quantitative_table}. Among the $12$ successfully launched model extraction attack, $10$ have successful watermark verification in the surrogate models, showing promising watermark transferability even without touching $\mathbb{M}_\theta$. The $2$ failure cases belong to the set of Experiments $11$--$14$ in which only the choice of $m$ is different. Despite the insensitivity to $m$ when verifying watermark in the original model, the verification in surrogate models is more sensitive to $m$. Furthermore, we emphasize that the successful verification in these surrogate models stems from a broader form of manifold transfer (or behavioral cloning) rather than architectural similarity. As shown in Table~\ref{tab:quantitative_table}, our experimental pairs exhibit massive architectural and parameter gaps. For instance, the Transformer-based OneRestore is stolen by a CNN-based U-Net (Experiment $9$), and the highly lightweight AODnet ($0.002$M) is stolen by a massive U-Net (Experiment $1$--$6$). By minimizing the functional discrepancy, the surrogate model inadvertently learns the latent watermark trajectory mapped by $\mathbb{M}_\theta$, allowing the watermark to survive regardless of the surrogate's internal structural design.

To provide more insights, we take a close look at $6$ experiments in Table \ref{tab:quantitative_table} and plot their loss and $\mathtt{NCC}$ values versus epoch during surrogate training, and the results are shown in Fig.~\ref{fig:line_chart_surrogate}, where the $\mathtt{NCC}$ values from the correct and incorrect key during the training of $\mathbb{G}_\gamma$ and $\mathbb{D}_\delta$ are also presented for comparison. The blue loss curves converge in all cases, indicating successful surrogate training. It can be seen from the green diamond curves that except for the failure cases in Fig.~\ref{fig:line_chart_surrogate} (d) and (e), the watermarking channel can be learned during the surrogate training in all cases, verifying the transferability of the watermark in these experiments.

Regarding the failure cases, for Experiment 12 (GAN-Binary), the problem lies in the metric calculation combined with the decoder's bias. Because a binary pattern possesses extreme pixel variance, the decoder projects invalid random queries into a high-contrast binary subspace, elevating the baseline $\mathtt{NCC}$ for both the original (red curve) and surrogate (pink curve) models, thus violating the $\mathtt{NCCD}$ criterion. Conversely, a zero-variance pure color (Exp. 14) mathematically suppresses the correlation of random noise to zero. For Experiment 13 (GAN-Pepper), the complex pepper watermark prevents both the correct and incorrect keys from being transferred to the surrogate model.

These observations provide practical guidance for watermark selection. For optimal robustness, we recommend selecting watermarks with moderate complexity and distinct semantic features (e.g., logos or text like ``COPYRIGHT"). Extreme structural simplicity (e.g., binary patterns) risks high false-positive baselines due to variance biases, whereas excessive complexity (e.g., complex images) hinders successful manifold transfer during surrogate training.

Additionally, to justify the empirical threshold of $0.5$ used in Eq. (\ref{eq:verify_s}), we analyze the sensitivity of the $\mathtt{NCCD}$ metric. Unlike the $\mathtt{NCC}$ score, which can drop depending on the surrogate model's extraction quality, $\mathtt{NCCD}$ operates as a differential metric that cancels out baseline degradations. As observed in the surrogate training curves in Fig.~\ref{fig:line_chart_surrogate}, while the $\mathtt{NCC}$ values fluctuate during surrogate training, the performance margin between the correct and incorrect keys remains remarkably stable. Furthermore, across the diverse architectures in Table~\ref{tab:quantitative_table}, the $\mathtt{NCCD}$ for valid keys consistently ranges between $0.60$ and $0.88$. Thus, $0.5$ serves as a highly robust, architecture-agnostic decision boundary that cleanly separates positive and negative ownership verifications.

\subsubsection{Key Security Against Brute Attack}
\label{subsubsec:exp_key_security}
We now test the key security of $\mathtt{ShadowMark}$ against brute force key ambiguity attacks. We assume the length and distribution of $\mathsf{k}$ is publicly known and set the number of random guesses to $1$ million times in each experiment, and the results are shown in the last column of Table \ref{tab:quantitative_table}. It can be seen that for all $16$ experiments, $\mathtt{SR}_\text{A} = 0$. To provide more insights, we are interested in measuring the theoretical key security. To achieve this, from the model owner's perspective, we define the relative distance between the true key and the guessed key by $\|\mathsf{k} - \tilde{\mathsf{k}}\|_2/\|\mathsf{k}\|_2$ and generate $\tilde{\mathsf{k}}$ randomly under different distance settings. For each relative distance, we calculate the ratio between discovered ambiguous keys and total number of guesses, denoted by $\mathtt{SR}_\text{O}$. The theoretical ambiguity test results are presented in Fig.~\ref{fig:radius_security}, which reveal the upper bounds of brute force key ambiguity attacks. It can be seen that the theoretical ambiguity is sensitive to the protected model, while in our experiments, the key for OneRestore is the most secure, followed by GAN and CGAN. In general, we notice that the key for $\mathtt{ShadowMark}$ has nontrivial ambiguous regions, and a rigorous quantification of these bounds is further discussed in Section \ref{sec:disscussion}. 

\subsubsection{PDJS Performance}
The results in Fig.~\ref{fig:PDJS_performance} (b) demonstrate that PDJS achieves significantly lower average service completion times compared to FIFO as client (model owner) concurrency increases under UCR-limited scenario. By effectively mitigating head-of-line blocking caused by heterogeneous workloads, PDJS accelerates overall system throughput without increasing the maximum completion time for the slowest owners. The efficiency gain stems from the payload-balancing capability visualized in Fig.~\ref{fig:PDJS_performance} (c). PDJS maintains near-optimal resource utilization ($\approx 1.0$) during active phases. Its payload-balancing mechanism operates akin to a ``sands and stones'' analogy: just as sand fills the gaps around large stones in a jar, the scheduler prioritizes tasks with low communication demands (sands) when the channel is congested by heavy uploads (stones), thereby utilizing the otherwise idle server and downlink resources.


Finally, the real-time profiles in Fig.~\ref{fig:PDJS_performance} (d)–(g) validate the efficiency and reliability of PDJS across diverse bottlenecks (UCR, SCR, and DCR) and the resource-balanced scenario. Unlike FIFO, where a single bottleneck causes head-of-line blocking that halts the entire pipeline, PDJS dynamically prioritizes tasks based on the proportion disparity distance $d_z$, enabling tasks utilizing non-limited resources to be processed and released rapidly, thereby alleviating congestion pressure on critical bottlenecks and accelerating overall service completion.

\begin{figure}[!t]
    \centering
    \includegraphics[width=1.0\columnwidth]{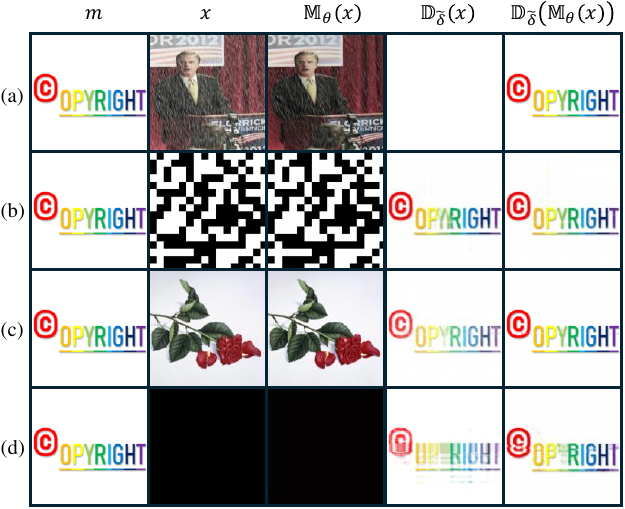}
    \caption{Ablation study results on key encoder $\mathbb{G}_\gamma$, where image restoration using OneRestore \cite{guo2024onerestore_Victim} is considered as an example. (a) $x \in \mathcal{X}$. (b)--(d) $x \notin \mathcal{X}$.}
    \label{fig:ablation_study}
    \vspace{-4pt}
\end{figure}

\subsection{Ablation Study}
The novel DNN components in $\mathtt{ShadowMark}$ are the key encoder $\mathbb{G}_\gamma$ and watermark decoder $\mathbb{D}_\delta$. Since $\mathbb{D}_\delta$ widely exists in DNN watermarking and is an essential component for watermark extraction, our ablation study mainly focuses on the necessity of incorporating $\mathbb{G}_\gamma$. Our experimental results reveal that the implementation without $\mathbb{G}_\gamma$ suffers from false verifications, meaning that $m$ can be extracted from non-watermarked images or equivalently images not processed by $\mathbb{M}_\theta$. This problem is demonstrated in Fig.~\ref{fig:ablation_study}, in which Fig.~\ref{fig:ablation_study} (a) shows that the in-distribution image after processed by $\mathbb{M}_\theta$ contains the watermark (and it does not contain watermark before being processed), while false verification occurs in Fig.~\ref{fig:ablation_study} (b)--(d), as can be seen from the $\mathbb{D}_{\tilde{\delta}}(x)$ column, in which images not processed by $\mathbb{M}_\theta$ can still have watermark extracted. Therefore, it verifies the necessity of $\mathbb{G}_\gamma$ that enables non-intrusive watermarking in the box-free sense. 

\section{Discussions and Future Work}
\label{sec:disscussion}

\subsection{Key Ambiguity and Security Boundaries}
As indicated in Section~\ref{subsubsec:exp_key_security}, while $\mathtt{ShadowMark}$ exhibits high resistance to brute-force attacks, theoretical analysis suggests the existence of ambiguous regions in the key space $\mathcal{K}$. This phenomenon stems from the intrinsic property of the high-dimensional manifold of deep neural networks. Since $\mathtt{ShadowMark}$ does not modify the model parameters $\theta$, the watermarking mapping $\mathbb{D}_\delta(\mathbb{M}_\theta(\mathbb{G}_\gamma(\mathsf{k}))) \to m$ essentially searches for a latent trajectory within the existing manifold of $\mathbb{M}_\theta$. Given the vast dimensionality of the input space $\mathcal{X}$, it is statistically possible that multiple distinct inputs (or keys) could trigger similar activation patterns leading to the watermark $m$.

To rigorously quantify this ambiguity and establish a security bound, calculating the exact analytical volume of these regions for arbitrary DNNs is intractable due to their highly non-linear nature. However, a probabilistic security bound can be derived from high-dimensional geometry. Assuming the key space $\mathcal{K}$ has a dimensionality of $d$, an ambiguity region can be mathematically modeled as an $\epsilon$-neighborhood around the true key $\mathsf{k}$. The volume of a $d$-dimensional hypersphere is proportional to its radius raised to the power of $d$. Consequently, the probability $P$ of a randomly sampled key falling within the $\epsilon$-radius ambiguity region in a normalized key space of radius $R$ scales exponentially: $P \propto (\epsilon / R)^d$. Given the high dimensionality of $d$, this probability decays exponentially toward zero. This theoretical derivation aligns well with our empirical findings in Fig.~\ref{fig:radius_security}, where the success rate of a brute-force attack drops precipitously as the relative distance increases. Crucially, because the true key is absolutely secret, a practical attacker is forced to rely entirely on global random guessing within the known key distribution. They have no reference point to perform localized guessing. The relative distance experiments (Fig. \ref{fig:radius_security}) serve purely as an oracle-based theoretical stress test to map the manifold boundaries, rather than a practical attack vector.

Furthermore, this theoretical ambiguity does not compromise practical security in the NWaaS context for two reasons. First, the specific side-channel is constrained by the trained key generator $\mathbb{G}_\gamma$. The valid keys must follow a specific distribution learned by $\mathbb{G}_\gamma$, making random noise attacks (as demonstrated in our experiments) ineffective. Second, in a real-world MLaaS API, the provider can easily implement rate limiting and query pattern analysis to detect and block brute-force attempts that aim to traverse the manifold for collisions. While the probabilistic bound proves the practical security of $\mathtt{ShadowMark}$, finding a closed-form analytical bound for specific model manifolds is still a challenge, which we highlight as a key area for future exploration.


\subsection{Vulnerability to Adaptive Attacks on the Side-Channel}
\label{subsec:adaptive_attacks}
Although NWaaS achieves several desirable properties due to its non-intrusive nature and decoupled verification design, its side-channel still relies on the trigger query $\mathbb{G}_\gamma(\mathsf{k})$, which exhibits distinct, out-of-distribution (OOD) features. Consequently, after training and deploying the surrogate model, an adversary is able to evade ownership verification at their service end. Specifically, the adversary can deploy query anomaly filters to reject non-natural inputs, or apply routine preprocessings to disrupt the fragile watermark manifolds of the trigger queries before they reach the model. To mitigate these adaptive threats, future research will focus on developing semantic-preserving key generators that seamlessly hide watermark triggers within natural image distributions, rendering verification queries indistinguishable from standard user traffic.

\subsection{Extension to Large Language Models (LLMs)} The current implementation of NWaaS focuses on X-to-Image models characterized by continuous and high-dimensional output spaces. Extending this framework to Large Language Models (LLMs) presents unique challenges due to the discrete nature of text generation. Specifically, the conversion from continuous logits to discrete tokens (e.g., via argmax or sampling) is mathematically non-differentiable. This breaks the gradient flow necessary for establishing the end-to-end side-channel of $\mathtt{ShadowMark}$ without modifying the target model's parameters. While recent LLM watermarking methods, such as PLMmark~\cite{li2023plmmark} and NSmark~\cite{zhaonsmark}, have achieved robust verification, they fundamentally rely on an intrusive paradigm that necessitates fine-tuning or updating the model parameters to alter the feature space or null space. 

Nevertheless, the core philosophy of NWaaS regarding non-intrusiveness remains applicable. To bypass the differentiability bottleneck, the watermarking signal could be embedded within the continuous probability distribution of next-token predictions (logits) or the hidden states of specific layers instead of the final discrete output. Alternatively, continuous relaxation techniques, such as the Gumbel-Softmax trick~\cite{jang2017categorical}, could be employed to approximate differentiability during training. Furthermore, adapting collaborative partitioning and PDJS scheduling for the massive parameter sets of LLMs, such as partitioning transformer blocks, to balance inference latency and IP protection is a promising direction for future exploration.

\section{Conclusion}
In this paper, we have presented NWaaS, a novel Non-intrusive WaaS designed to address the critical challenges of intrusiveness, privacy leakage, and inefficiency in existing IP protection paradigms. At the core of our system, we have introduced $\mathtt{ShadowMark}$, an algorithm that establishes a verifiable side-channel for ownership verification. By decoupling the watermark embedding from the target model's normal functionality, $\mathtt{ShadowMark}$ ensures zero performance degradation and eliminates the need for accessing original training data for watermarking. Building upon this non-intrusive property, NWaaS implements a flexible collaborative partitioning mechanism, allowing model owners to submit self-defined partial non-sensitive layers to the cloud server while retaining IP-critical layers locally, thereby resolving the fundamental trust issues associated with third-party service providers. To further ensure scalability, resource efficiency, and low latency under high concurrency scenario, we have proposed PDJS, balancing the heterogeneous demands of uplink, server-side computation, and downlink resources. Extensive experiments across diverse settings (covering different X-to-Image tasks, diverse datasets, model architectures, and resource scenarios) have demonstrated that NWaaS provides robust ownership verification and model stealing tracing against model extraction attack, while successfully reconciling the conflicting goals of high-fidelity model protection, rigorous owner privacy, and operational efficiency, thereby offering a trustworthy solution for the evolving MLaaS system.

\bibliographystyle{IEEEtran}
\bibliography{ref}

\begin{IEEEbiography}[{\includegraphics[width=1in,height=1.25in,clip,keepaspectratio]{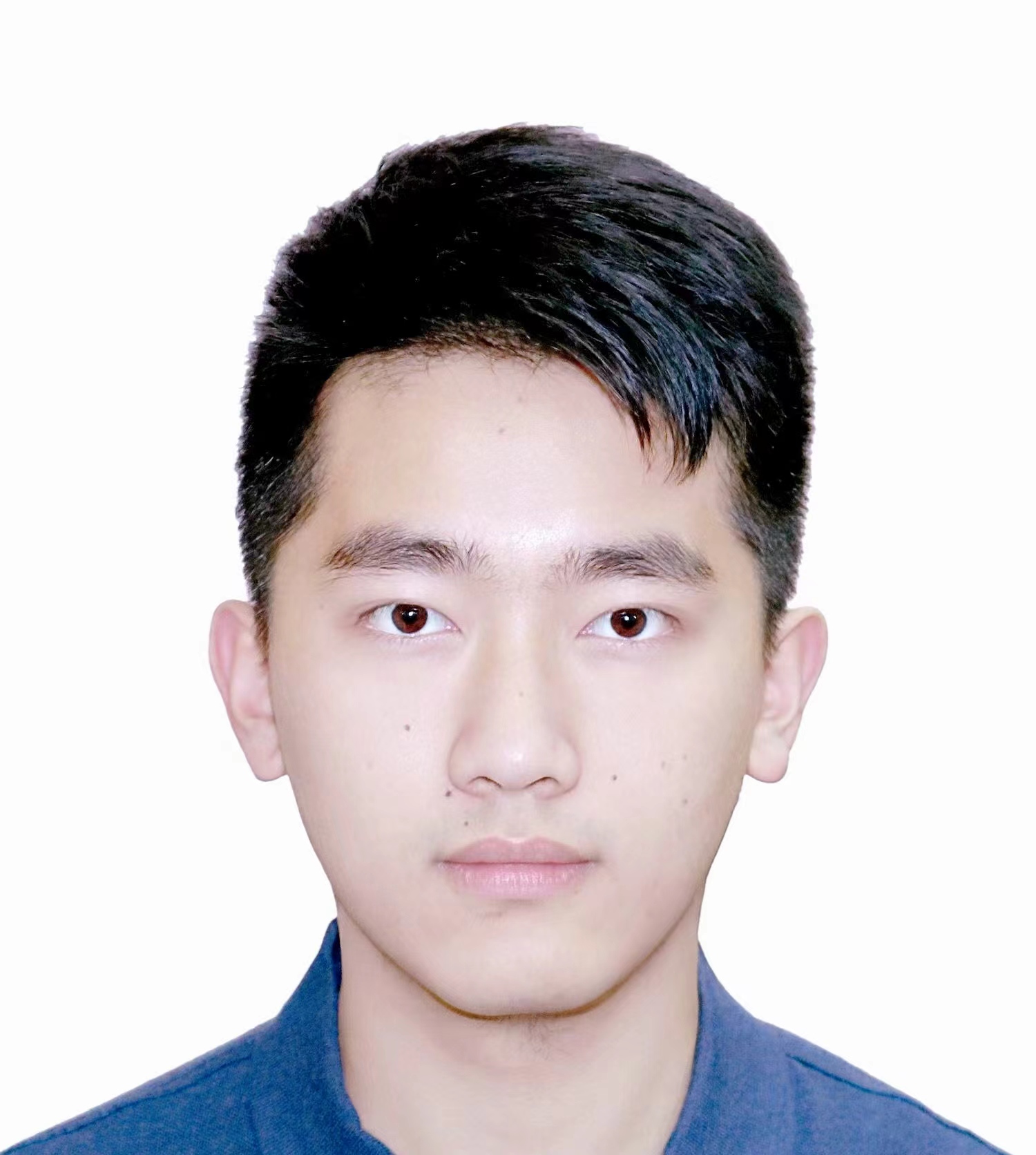}}]{Haonan An} (Graduate Student Member, IEEE)
received his B.Eng. degree in Telecommunication Engineering from Huazhong University of Science and Technology, Wuhan, China, in 2023. He obtained his MS degree in Signal Processing from the School of Electrical and Electronic Engineering, Nanyang Technological University, Singapore, in 2024. He is currently pursuing his Ph.D. at the City University of Hong Kong. His research interests include IP protection, model watermarking, and AIGC forensics.
\end{IEEEbiography}

\begin{IEEEbiography}[{\includegraphics[width=1in,height=1.25in,clip,keepaspectratio]{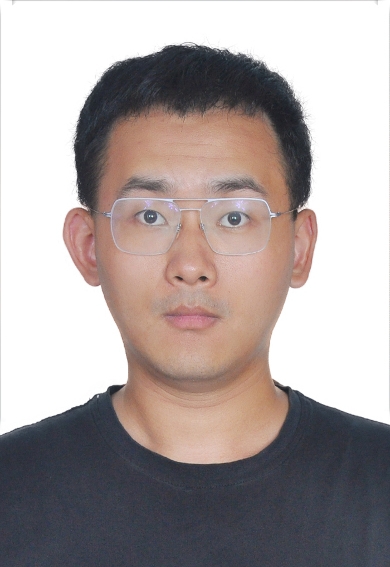}}]{Qianyao Ren} is currently pursuing his Ph.D. degree in the Department of Computer Science, City University of Hong Kong. He received his B.S. and M.S. degrees from the school of Electronics and Information Engineering, Harbin Institute of Technology in 2015 and 2017. His research interests include massive MIMO and AI-enhanced wireless communication.

\end{IEEEbiography}

\begin{IEEEbiography}[{\includegraphics[width=1in,height=1.25in,clip,keepaspectratio]{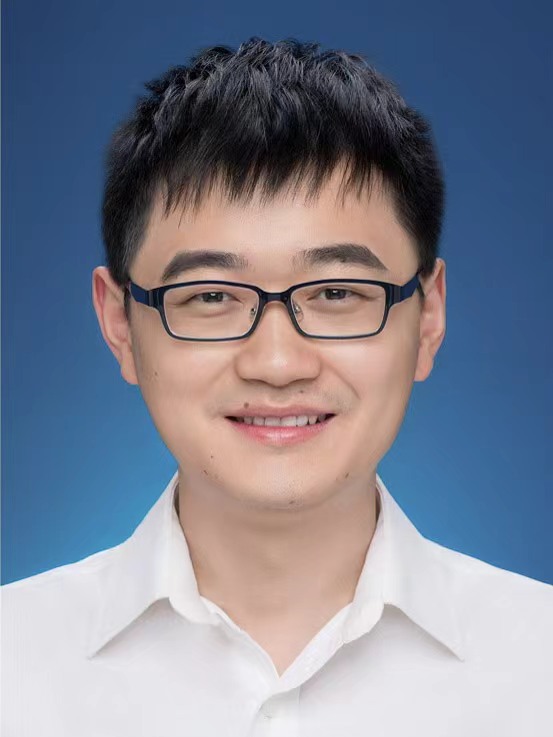}}]{Guang Hua} received the B.Eng. degree (Communication Engineering) from Wuhan University (WHU), China, in 2009, and the Ph.D. degree (Information Engineering) from Nanyang Technological University (NTU), Singapore, in 2014. He was a Research Fellow with the School of Electrical and Electronic Engineering, NTU (2015–2016), an Associate Professor with the School of Electronic Information, WHU (2017–2022), an International Scholar Exchange Fellow at Yonsei University, sponsored by the CHEY Institute for Advanced Studies, South Korea (2020–2021), a Scientist (2013–2015) and a Senior Scientist (2022–2023) with the Institute for Infocomm Research (I2R), A*STAR, Singapore. He is currently an Associate Professor with the Infocomm Technology (ICT) Cluster, Singapore Institute of Technology (SIT). He was the PI for two China’s NSFC projects, one Singapore MOE project, and several industry projects. His research interests include AI security, media security, and statistical signal processing, for which he has publications in IEEE TPAMI, TIFS, TDSC, TSP, TASLP, TNNLS, TCSVT, CVPR, AAAI, ICASSP, WIFS, NDSS, etc. He holds a Singapore patent (licensed) and a Chinese patent (transferred) on media forensics. He is an IEEE senior member, Associate Editor for IEEE TIFS, and the Senior Area Editor for IEEE SPL.
\end{IEEEbiography}

\begin{IEEEbiography}[{\includegraphics[width=1in,height=1.25in,clip,keepaspectratio]{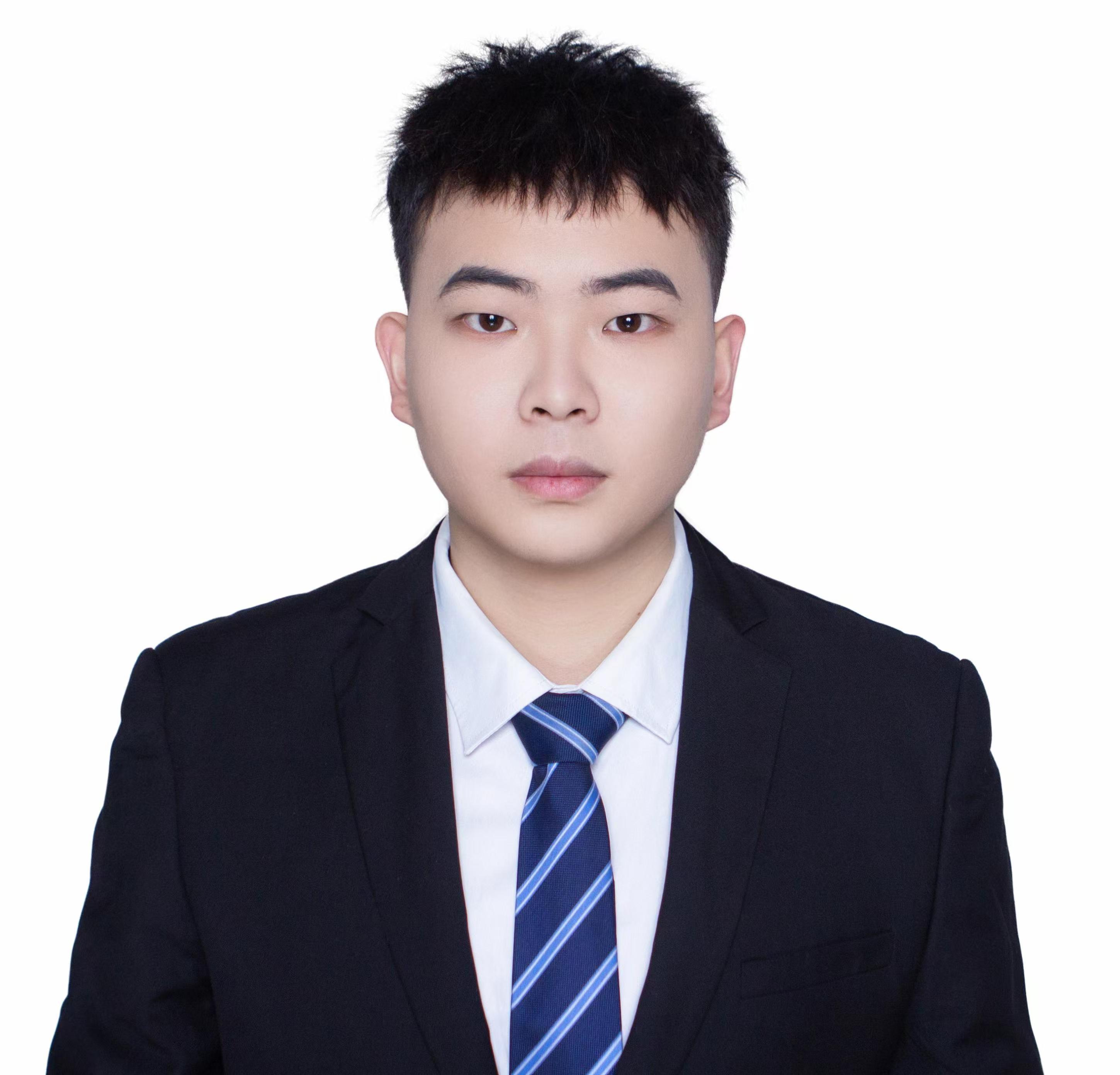}}]{Tao Li} obtained his B.Eng. degree in Electronic and Information Engineering from the China University of Geosciences (Wuhan) in 2023, and completed his MSc degree in Electrical and Electronic Engineering at the University of Hong Kong in 2024. He is currently working towards his Ph.D. degree in the Department of Electrical and Electronic Engineering at the University of Hong Kong. His research interests include mobile computing, split and federated learning, and distributed deployment of large language models.
\end{IEEEbiography}

\begin{IEEEbiography}
[{\includegraphics[width=1in,height=1.25in,clip,keepaspectratio]{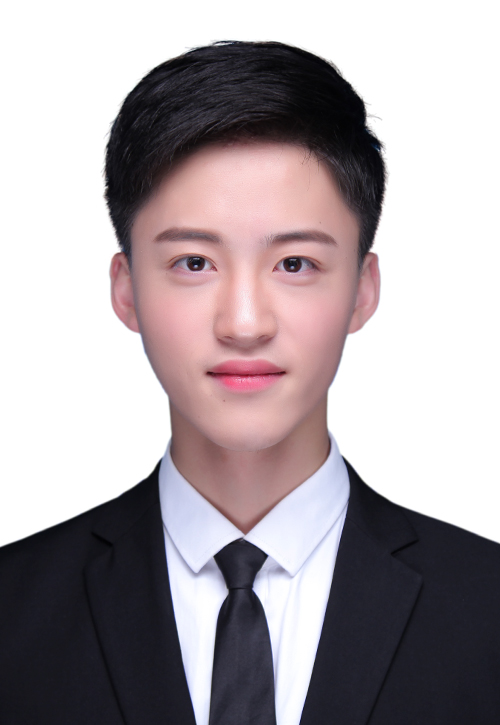}}]{Yu Guo}
received a B.Sc. degree in naval architecture and marine engineering from the School of Transportation, Wuhan University of Technology, Wuhan, China, in 2021. He is pursuing a Ph.D degree in traffic information engineering and control at the School of Navigation, Wuhan University of Technology, and he is currently a research assistant at WINET lab, City University of Hong Kong. His research interests include computer vision, machine learning, and intelligent transportation systems.
\end{IEEEbiography}

\begin{IEEEbiography}[{\includegraphics[width=1in,height=1.25in,clip,keepaspectratio]{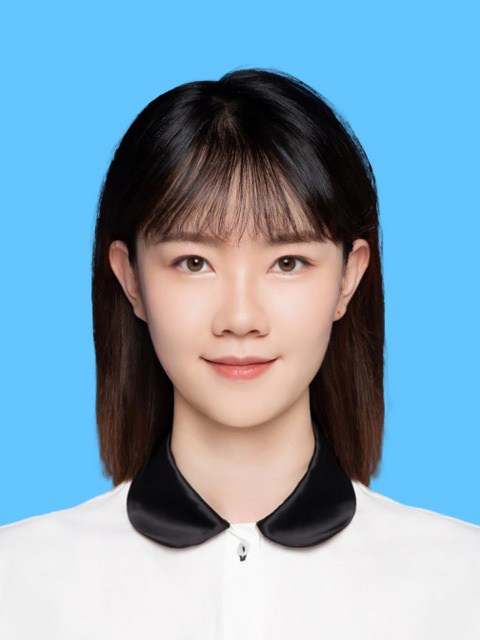}}]{Yanan Ma} (Graduate Student Member, IEEE)
received the B.Eng. degree (Hons.) in Electronic Information Engineering (English Intensive) and the M.Eng. degree (Hons.) in Information and Communication Engineering from the Dalian University of Technology, Dalian, China, in 2020 and 2023, respectively. She is currently pursuing the Ph.D. degree in the Department of Computer Science at the City University of Hong Kong. She received the IEEE GLOBECOM Best Paper Award in 2025. Her research interests are focused on edge intelligence,  wireless communication and networking, and machine learning.
\end{IEEEbiography}

\begin{IEEEbiography}[{\includegraphics[width=1in,height=1.25in,clip,keepaspectratio]{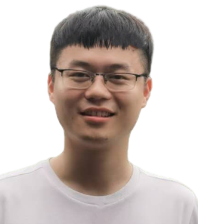}}]{Hangcheng Cao}
is currently a postdoctoral fellow in the Department of Computer Science, City University of Hong Kong, China. He obtained the Ph.D. degree in the College of Computer Science and Electronic Engineer, Hunan University, China, in 2023. He studied as a joint PhD student in the School of Computer Science and Engineering, Nanyang Technological University, Singapore, in 2022. He has published papers in IEEE S\&P, ACM Ubicomp/IMWUT, IEEE INFOCOM, IEEE ICDCS, IEEE TMC, ACM ToSN, IEEE Communications Magazine, Information Fusion, IEEE IoT-J, IEEE JSAS, etc. His research interests lie in the area of IoT security.\end{IEEEbiography}

\begin{IEEEbiography}[{\includegraphics[width=1in,height=1.25in,clip,keepaspectratio]{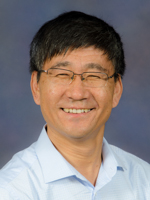}}]{Yuguang Fang}
received MS from Qufu Normal University, China, PhD from Case Western Reserve University, USA, and PhD from Boston University, USA, in 1987, 1994, and 1997, respectively. He joined the Department of Electrical and Computer Engineering at University of Florida in 2000 as an assistant professor, then was promoted to associate professor, full professor, and distinguished professor, in 2003, 2005, and 2019, respectively. Since 2022, he has been a Global STEM Scholar and Chair Professor with Department of Computer Science, City University of Hong Kong. 

He received many awards including US NSF CAREER Award, US ONR Young Investigator Award, 2018 IEEE Vehicular Technology Outstanding Service Award, and IEEE Communications Society awards (AHSN Technical Achievement Award, CISTC Technical Recognition Award, and WTC Recognition Award). He was Editor-in-Chief of IEEE Transactions on Vehicular Technology and IEEE Wireless Communications. He is a fellow of ACM, IEEE, and AAAS.  
\end{IEEEbiography}

\vfill

\end{document}